\documentclass[fleqn,usenatbib]{mnras}

\usepackage[T1]{fontenc}
\usepackage{ae,aecompl}

\usepackage{graphicx}	
\usepackage{amsmath}	
\usepackage{amssymb}	

\title[Bubble mapping with the SKA]{Bubble mapping with the Square Kilometer Array -- I.\\ Detecting galaxies with Euclid, JWST, WFIRST and ELT within ionized bubbles in the intergalactic medium at $z>6$}

\author[Zackrisson et al.]{Erik Zackrisson,$^{1}$\thanks{E-mail: erik.zackrisson@physics.uu.se} 
Suman Majumdar,$^{2}$ 
Rajesh Mondal,$^{3}$
Christian Binggeli,$^{1}$\newauthor
Martin Sahl\'en,$^{1}$
Tirthankar Roy Choudhury,$^{4}$
Benedetta Ciardi,$^{5}$
Abhirup Datta,$^{2}$\newauthor
Kanan K. Datta,$^{6}$
Pratika Dayal,$^{7}$
Andrea Ferrara,$^{8,9}$ 
Sambit K. Giri,$^{10}$\newauthor
Umberto Maio,$^{11}$
Sangeeta Malhotra,$^{12,13}$
Garrelt Mellema,$^{10}$
Andrei Mesinger,$^{8}$\newauthor 
James Rhoads,$^{12,13}$
Claes-Erik Rydberg,$^{14}$
Ikkoh Shimizu $^{15}$ 
\\
$^{1}$Observational Astrophysics, Department of Physics and Astronomy, Uppsala University, Box 516, SE-751 20 Uppsala, Sweden\\
$^{2}$Discipline of Astronomy, Astrophysics and Space Engineering, Indian Institute of Technology Indore, Simrol, Indore 453552, India\\
$^{3}$Astronomy Centre, Department of Physics and Astronomy, University of Sussex, Brighton  BN1 9QH, UK\\
$^{4}$National Centre for Radio Astrophysics, TIFR, Post Bag 3, Ganeshkhind, Pune 411007, India\\
$^{5}$Max-Planck-Institut f\"{u}r Astrophysik, Karl-Schwarzschild-Str. 1 D-85748 Garching, Germany\\
$^{6}$Department of Physics, Presidency University, 86/1 College Street, Kolkata 700073, India\\
$^{7}$Kapteyn Astronomical Institute, University of Groningen, P.O. Box 800, 9700 AV Groningen, The Netherlands \\
$^{8}$Scuola Normale Superiore, Piazza dei Cavalieri 7, I-56126, Pisa, Italy\\
$^{9}$Kavli IPMU, The University of Tokyo, 5-1-5 Kashiwanoha, Kashiwa 277-8583, Japan\\
$^{10}$Department of Astronomy, Stockholm University, Oskar Klein Center, AlbaNova, Stockholm SE 106 91, Sweden\\
$^{11}$Max Planck Institute for Astrophysics, Karl-Schwarzschild-Str. 1, 85741 Garching, Germany\\
$^{12}$Arizona State University, School of Earth and Space Exploration, Tempe, AZ 85287, USA\\
$^{13}$NASAs Goddard Space Flight Center, Astrophysics Science Division, Code 660, Greenbelt MD 20771, USA\\
$^{14}$Universit\"{a} Heidelberg, Zentrum f\"{u}r Astronomie, Institut f\:{u}r Theoretische Astrophysik, Albert-Ueberly-Str. 2, 69120 Heidelberg, Germany\\
$^{15}$Department of Earth and Space Science, Osaka University, 1-1 Machikaneyama, Toyonaka, Osaka 560-0043, Japan
}

\date{Accepted XXX. Received YYY; in original form ZZZ}

\pubyear{2019}

\begin{document}
\label{firstpage}
\pagerange{\pageref{firstpage}--\pageref{lastpage}}
\maketitle

\begin{abstract}
The Square Kilometer Array is expected to provide the first tomographic observations of the neutral intergalactic medium at redshifts $z>6$ and pinpoint the locations of individual ionized bubbles during early stages of cosmic reionization. In scenarios where star-forming galaxies provide most of the ionizing photons required for cosmic reionization, one expects the first ionized bubbles to be centered on overdensities of such galaxies. Here, we model the properties of galaxy populations within isolated, ionized bubbles that SKA-1 should be able to resolve at $z\approx 7$--10, and explore the prospects for galaxy counts within such structures with various upcoming near-infrared telescopes. We find that, for the bubbles that are within reach of SKA-1 tomography, the bubble volume is closely tied to the number of ionizing photons that have escaped from the galaxies within. In the case of galaxy-dominated reionization, galaxies are expected to turn up above the spectroscopic detection threshold of JWST and ELT in even the smallest resolvable bubbles at $z\leq 10$. The prospects of detecting galaxies within these structures in purely photometric surveys with Euclid, WFIRST, JWST or ELT are also discussed. While spectroscopy is preferable towards the end of reionization to provide a robust sample of bubble members, multiband imaging may be a competitive option for bubbles at $z\approx 10$, due to the very small number of line-of-sight interlopers expected at these redshifts.
\end{abstract}




\section{Introduction}
\label{intro}
In the currently favoured view of galaxy-dominated reionization, large ionized bubbles in the intergalactic medium (IGM) will first appear around overdensities of galaxies, progressively grow and finally coalesce \citep[for recent reviews, see][]{loeb13,Barkana16,Mesinger16,dayal2018}. Upcoming observations of the redshifted 21 cm signal from the neutral IGM will open a new window on this process, and existing constraints from the high-redshift galaxy luminosity function, from the cosmic microwave background radiation and from quasar absorption systems can be used to forecast the viable range of 21 cm signals from neutral hydrogen in the reionization epoch \citep[e.g.][]{Kulkarni16,Hassan17,Mirocha17,Greig17a}. 

While current interferometers are limited to detecting the 21 cm signal in a statistical sense (for instance the 21 cm power spectrum), phase one of the Square Kilometer Array (hereafter SKA-1) will be able to resolve physical scales down to 5--10 comoving Mpc in the plane of the sky and corresponding physical scales along the line of sight (frequency) direction at $z\approx 6$--10 \citep{Mellema15,Wyithe15,Datta16,Mondal18,Mondal18a}. This will for the first time allow tomography (3-dimensional imaging) of the 21 cm signal. 

It is already well established that 21 cm data correlated with galaxy surveys can provide powerful constraints on reionization scenarios \citep{WyitheLoeb07,Lidz09,Wiersma13,Park14,Vrbanec16,Hasegawa16,Sobacchi16,Hutter17,Hutter18}. However, most of the studies in this field have focused on the prospects of wide-field (and therefore comparatively shallow) galaxy surveys, whereas relatively little effort has been devoted to the prospects of deep, small-field surveys that focus on uncovering the galaxy content of individual ionized bubbles \citep[but see][for discussions on how to combine MWA/HERA/SKA data with Wide-Field Infrared Survey Telescope (WFIRST) and James Webb Space Telescope (JWST) data this way]{Beardsley15,Geil17}.

The sharpness of the 21 cm profile at the edge of the ionizing region can provide information on the distribution of ionization sources within (e.g. a single quasar vs. a spatially extended group of galaxies; \citealt{wyithe05, datta07, datta08,datta12,Datta16, majumdar11,majumdar12,malloy13,Kakiichi17, Giri18}) and also on the relative contribution of X-ray and ultraviolet photons within the bubble (quasar/mini-quasar/high-mass X-ray binaries vs. young stars; e.g. \citealt{Tozzi00,Wyithe07,Pacucci14,Ghara16,Kakiichi17}). 

Here, we will use relatively simple simulations to explore what one can hope to learn by combining an SKA-1 measurement of the dimensions of an individual ionized IGM bubble with a photometric/spectroscopic galaxy survey of its content using upcoming telescopes like the JWST, Euclid, WFIRST or the Extremely Large Telescope (ELT). By mapping the galaxy content of individual, relatively isolated bubbles, it may be possible to assess the ionizing photon budget within these regions and constrain the role of the galaxies detected inside, in a more direct way than what can currently be done for the photon budget of ionized regions surrounding $z\gtrsim 6$ Lyman-$\alpha$ emitters \citep[e.g.][]{Bagley17,Yajima17,Castellano18}, as both the total ionizing photon budget and the contribution from galaxies not exhibiting detectable Lyman-$\alpha$ emission within such regions tends to remain ambiguous. 

What galaxies are expected inside regions of the Universe that reionize early? Depending on the redshift, size and isolation of such structures, these regions may be highly biased and could in principle contain galaxies with properties that deviate substantially from those in the average galaxy population at the same redshift. Throughout this paper, we will however adopt the conservative assumption that the galaxies clustered within 21 cm bubbles exhibit higher number densities but properties otherwise identical to those in the field population at the same epoch. This zeroth-order estimate can then serve as a benchmark for more detailed simulations in future efforts.

In Section~\ref{budget}, we explain how mapping the galaxy populations within ionized regions of the IGM at $z\gtrsim 7$ can provide constraints on the role of galaxies in the emergence of these structures. Using semi-numerical simulations of galaxy-dominated reionization, we in Section~\ref{sizes} predict the relation and scatter between the number of ionizing photons emitted from galaxies within a bubble and the resulting volume of that structure, as a function of redshift. The detection limits for galaxies within these structures are explored in Section~\ref{galaxies}. In Section~\ref{discussion}, a number of simplifications adopted in this work are discussed. We also comment on the prospects of using populations of bubble galaxies to constrain early assembly/environmental bias and to place combined constraints on the luminosity function of bubble galaxies and on the time-integrated mean escape fraction of ionizing photons from these objects. Section~\ref{summary} summarizes our findings.

\section{The photon budget of ionized bubbles}
\label{budget}
Considering a spherical ionized region of comoving radius $r$ and volume $V_\mathrm{ion} =  (4/3)\pi r^3$ and ignoring the effect of recombinations inside it, the relationship between the comoving ionized volume and the total number of ionizing photons $N_\mathrm{ion,tot}$ that has ever been emitted into the IGM in this region can be expressed as:
\begin{equation}
V_\mathrm{ion}\approx \frac{N_\mathrm{ion,tot}} {\langle n_\mathrm{H} \rangle},
\label{Vbubble_eq2}
\end{equation}
where $\langle n_\mathrm{H} \rangle$ is the average comoving number density of hydrogen atoms in the IGM. In scenarios where star-forming galaxies provide the bulk of ionizing photons required for cosmic reionization, $N_\mathrm{ion,tot}$ corresponds to the total number of ionizing photons that have ever escaped from galaxies within the bubble.  

Eq.~(\ref{Vbubble_eq2}) suggests that, if SKA-1 is able to identify individual, highly ionized IGM bubbles and also estimate their volume ($V_\mathrm{ion}$), it may be possible to place a constraint on the integrated number of ionizing photons $N_\mathrm{ion,tot}$ emitted from galaxies within this structure. Formally, the $N_\mathrm{ion,tot}$ constraint inferred from eq.~(\ref{Vbubble_eq2}) will be a lower limit, since a greater number of ionizing photons will be required once recombinations are considered. However, this $N_\mathrm{ion,tot}$ estimate is, for reasonable assumptions on the IGM clumping factor, expected to be accurate to within a factor of a few \citep[e.g.][]{McQuinn07,Sobacchi14}. 

The number of ionizing photons {\it emitted} by the galaxy population into a specific region of the IGM is determined by the number of ionizing photons {\it produced}, modulo the escape fraction of these photons. Under the assumption of an invariant stellar initial mass function, the number of ionizing photons produced is, in turn, related to the total mass in stars produced in this region. However, neither the total mass in stars nor the total number of ionizing photons produced within a region are directly observable. All one can hope to detect is individual galaxies in the bright-end tail of the galaxy population within this volume. In what follows, we will explain how these quantities are related.

While IGM bubbles grow gradually, with galaxies in different mass and luminosity regimes contributing to $N_\mathrm{ion,tot}$ at different times, a constraint on $N_\mathrm{ion,tot}$ may nonetheless be converted into a rough estimate on the number of galaxies expected within that bubble at the epoch from which we detect the 21 cm signal. This is possible since the instantaneous, rest-frame 1500 \AA~ultraviolet (UV) luminosity $L_\mathrm{UV}$ (i.e. in the non-ionizing part of the UV; redshifted into the near-infrared at $z>6$), which traces recent star formation (over the past $\lesssim 10^8$ yr) within a galaxy is predicted to be correlated with the total stellar mass ever formed in that system and in all the progenitors that have merged into it. This stems from the generic simulation prediction that $z>6$ galaxies {\it on average} have star formation/accumulation rates that increase over time \citep[e.g.][]{Finlator11,Jaacks12,Dayal13,Shimizu14,Ma15,Zackrisson17}. 

The number of ionizing photons $N_{\mathrm{ion},i}$ emitted from a single galaxy $i$ into the IGM over its past star formation history up to the point in time when it is observed ($t_\mathrm{obs}$) can be expressed as:
\begin{equation}
N_{\mathrm{ion},i}=\int_0^{t_\mathrm{obs}} f_\mathrm{esc}(t) \dot{N}_\mathrm{ion}(t) \ \mathrm{d}t,
\label{Nion_i}
\end{equation}
where $\dot{N}_\mathrm{ion}(t)$ is the production rate of the number of ionizing photons in this galaxy at time $t$ and $f_\mathrm{esc}(t)$ describes the temporal evolution of the escape fraction of ionizing photons into the IGM. 

If we define $\langle f_\mathrm{esc} \rangle$ as the $N_\mathrm{ion}$-weighted mean $f_\mathrm{esc}$ over the past history of the galaxy, eq.~\ref{Nion_i} simplifies to:
\begin{equation}
N_{\mathrm{ion},i}=\langle f_\mathrm{esc} \rangle \int_0^{t_\mathrm{obs}} \dot{N}_\mathrm{ion}(t) \ \mathrm{d}t,
\label{Nion_i2}
\end{equation}

The total number of ionizing photons produced by a whole population of galaxies $N_\mathrm{ion,tot}$ in a volume $V_\mathrm{ion}$ can then be derived by integrating over galaxies of all UV luminosities, $L_\mathrm{UV}$: 
\begin{equation}
N_\mathrm{ion,tot}= \langle f_\mathrm{esc}\rangle \int_{L_\mathrm{min}}^{L_\mathrm{max}} N_\mathrm{ion,i}(L_\mathrm{UV}) \Phi(L_\mathrm{UV}) V_\mathrm{ion} \ \mathrm{d}L_\mathrm{UV},
\label{Niontot_eq}
\end{equation}
where $\Phi(L_\mathrm{UV})$ describes the luminosity function of galaxies in this ionized region (in units of galaxies per volume per unit 1500 \AA{} luminosity) -- which is going to have a much higher scaling than the galaxy luminosity function in the field. In this equation, we have for simplicity assumed that all galaxies have the same $\langle f_\mathrm{esc} \rangle$ (this assumption is relaxed in Section~\ref{discussion}) and that $N_\mathrm{ion,i}$ only depends on $L_\mathrm{UV}$. 

If we take $N_\mathrm{ion,i}$ to be known then eq.~(\ref{Niontot_eq}) indicates how an estimate on $N_\mathrm{ion,tot}$ (provided by SKA-1, via the bubble volume $V_\mathrm{ion}$ in eq.~\ref{Vbubble_eq2}) can be used to constrain the galaxy population ($\Phi(L_\mathrm{UV})$) within the bubble and the time-integrated escape fraction of ionizing photons $\langle f_\mathrm{esc}\rangle$ from the bubble galaxies. We may, for an individual ionized IGM bubble of a given size, conversely also provide a rough estimate on the number of galaxies that are expected to lie above some UV luminosity detection within this bubble, given an assumption on the relative shape or slope of the galaxy luminosity function within this bubble and on the likely value of $\langle f_\mathrm{esc} \rangle$. This allows us to assess the prospects of detecting bubble galaxies with some of the telescopes that are expected to be operational in the SKA-1 era, which we set out to do in the following sections.

In reality, $N_\mathrm{ion,i}$ will vary substantially from galaxy to galaxy of the same observed $L_\mathrm{UV}$ due to differences in star formation history, metallicity and dust attenuation, and we will in Sect~\ref{galaxies} use galaxy spectral energy distribution (SED) models coupled to galaxy simulations in an attempt to quantify the distribution of $N_\mathrm{ion,i}/L_\mathrm{UV}$, i.e. the total number of ionizing photons produced over the momentary UV luminosity, and its impact on the relation between galaxy counts and the ionizing photon budget. 

\section{The sizes of ionized bubbles}
\label{sizes}

\subsection{The smallest ionized bubbles detectable with SKA-1}
As shown by e.g. \citet[][]{Mellema15}, \citet{Wyithe15} and \citet{Datta16}, SKA-1 should be able to identify individual ionized IGM bubbles of angular diameters down to $\gtrsim 5\arcmin$\footnote{This is similar to the largest SKA1-LOW beam FWHM that minimizes the point spread function near-in sidelobe noise in full-track mode at 0.13-0.18 GHz} at $z\approx 6$--10, which corresponds to a spherical bubble radius of $\gtrsim 6$--7 cMpc or a spherical volume of $\gtrsim 1000$ cMpc$^3$. Ionized bubbles of this size are most readily detected using the matched filtering technique in the Fourier domain proposed by \citet{datta07,datta08}, \citet{majumdar11,majumdar12} and \citet{datta12,Datta16}. This technique optimally combines the complete 3-dimensional 21 cm signal from HI outside the bubble using a matched filter. This method also takes advantage of the fact that noise is uncorrelated in the Fourier domain, whereas it is correlated in the image domain, thereby resulting in a higher signal-to-noise ratio for a given bubble size than methods based on imaging \citep[e.g.][]{Mellema15,Kakiichi17,Giri18} or 1-dimensional 21 cm spectra \citep[e.g.][]{Geil17}. 

\subsection{Bubble simulations}
\label{bubble_section}
\begin{figure*}
\includegraphics[scale=0.85]{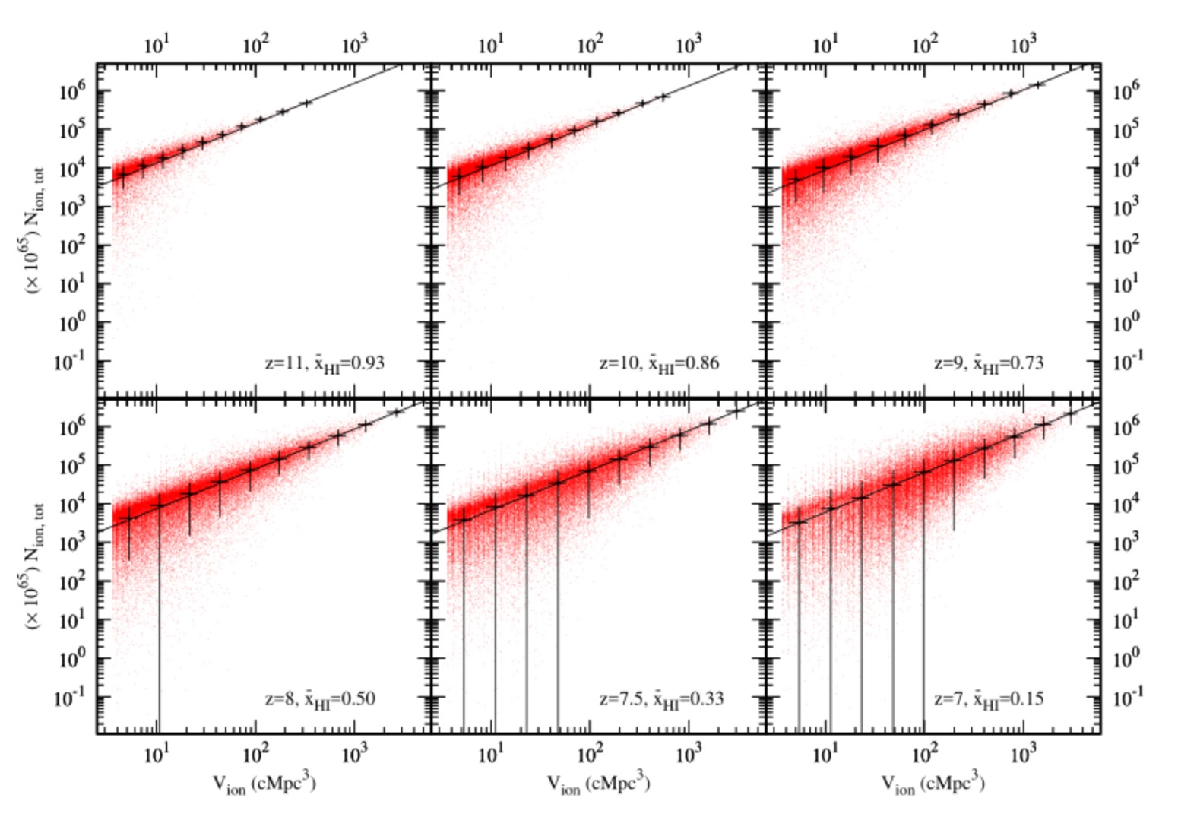}
\caption{Relation between the volumes of ionized IGM bubbles $V_\mathrm{ion}$ at $z=11$--7 and the number of ionizing photons $N_\mathrm{ion,tot}$ that have escaped from galaxies and into the IGM within each such region. Black crosses indicate the size of the 1$\sigma$ scatter in each bin. The different panels feature the neutral IGM fraction of our default reionization scenario along with the best-fitting $N_\mathrm{ion}-V_\mathrm{ion}$ relation at each redshift. For bubbles sufficiently large to be resolved by SKA-1 ($V_\mathrm{ion}\gtrsim 1000$ cMpc$^3$ in the case of spherical bubbles), the $1\sigma$ range in $N_\mathrm{ion}$ at fixed $V_\mathrm{ion}$ is always limited to a factor of $<4$.}
\label{bubble_sizes}
\end{figure*}
To explore how tightly coupled $V_\mathrm{ion}$ may be expected to be to $N_\mathrm{ion,tot}$, we use a set of semi-numerical simulations for reionization, which are identical to those presented by \citet{Mondal17}. These simulations involve three major steps: a) First we simulate the matter distribution at different redshifts using a publicly available particle mesh \textit{N}-body code\footnote{\url{https://github.com/rajeshmondal18/N-body}} and assume that hydrogen follows this underlying matter field; b) Next we identify collapsed structures in this matter distribution using a publicly available halo finder\footnote{\url{https://github.com/rajeshmondal18/FoF-Halo-finder}} based on the Friends-of-Friend (FoF) algorithm \citep{davis85}; c) We then assume a model for the sources of ionization hosted by these collapsed halos and generate an ionizing photon field using a publicly available semi-numerical code\footnote{\url{https://github.com/rajeshmondal18/ReionYuga}}. A general assumption in our ionizing source model is that the number of ionizing photons that are produced by these sources is proportional to their host halo mass $M_\mathrm{halo}$. We use the constant of proportionality $n_{\rm ion}$ (dimensionless) as a parameter for our simulations. This quantity (also known as the ionization efficiency) combines a number of reionization parameters e.g. the star formation efficiency, the fraction of ionizing photons escaping into the IGM, the number of ionizing photons per baryons produced etc. For a  detailed discussion on this we refer the readers to  Sec 2.3 of \citet{choudhury09}. Finally, we use this ionizing photon field and the matter density field under an excursion set formalism \citep{furlanetto04} to identify ionized regions within the hydrogen distribution \citep[e.g.][]{Zahn07,Mesinger07}. Our method of simulating the ionization fields during reionization is similar to that of \citet{choudhury09}, \citet{majumdar14}, \citet{Mondal15}, and \citet{mondal16}.

The $N$-body simulation that we use here has a comoving volume of $V = [215\, {\rm cMpc}]^3$, corresponding to $\sim 1.3^\circ$ on the sky for $7< z < 10$, with a $3072^3$ grid of spacing $0.07$ cMpc and a particle mass of $1.09 \times 10^8\, {\rm M_{\odot}}$. Thus the smallest dark matter halo that we can resolve is $1.09 \times 10^9\, {\rm M_{\odot}}$ (assuming a minimum of $10$ particles required to form a halo). Once we have identified the halos, we then map the matter and the ionizing photon density fields on a grid which is eight times coarser than our original $N$-body simulation resolution (i.e. on a $384^3$ grid). These coarser fields are then used to implement the excursion set formalism. We identify a grid point as neutral or ionized at a certain stage of reionization, by smoothing  and comparing the hydrogen density and the photon density fields using spheres of different radii starting from a minimum radius of $R_{{\rm min}}$ (the coarse grid spacing) to $R_{{\rm mfp}}$ (mean free path of the ionizing photons). A specific grid point is considered to be ionized if for any smoothing radius $R$ ($R_{{\rm min}} \leq R \leq R_{{\rm mfp}}$) the photon density exceeds the neutral hydrogen density at that grid point. For the simulation shown here we have used $n_{{\rm ion}} = 23.21$ and $R_{{\rm mfp}} =  20\, {\rm Mpc}$ (which is consistent with \citealt{songaila10}) at all redshifts. These values of the parameters ensure that reionization ends at $z \approx 6$ and we obtain a Thomson scattering optical depth $\tau = 0.057$, which is consistent with \citet{planck16a}. We have used the Planck+WP best fit values of cosmological parameters $\Omega_{\rm m}=0.3183$, $\Omega_{\rm \Lambda}=0.6817$, $\Omega_{\rm b}h^2=0.022032$, $h=0.6704$, $\sigma_8=0.8347$, and $n_{\rm s}=0.9619$ \citep{Planck14}.

Once ionization maps have been generated at a set of redshifts, we once again make use of a FoF algorithm on these gridded ionization maps to identify individual ionized regions. In this FoF algorithm, we identify any cell having a neutral fraction $x_{{\rm HI}} \leq 10^{-4}$ as ionized. 
This reionization model and numerical machinery results in several tens to hundreds of ionized IGM bubbles above the SKA-1 tomographic limit (volume $\gtrsim 1000$ cMpc$^3$) at $z\approx 7$--9 within our simulated volume. Rescaling these bubble counts to the volume covered by the planned 100 deg$^2$ deep SKA1-LOW survey \citep{Koopmans15} would result in $\approx 7\times 10^4$, $1\times 10^4$ and $1\times 10^3$ such bubbles per $\Delta(z)=1$ at $z\approx 7$, 8 and 9. Hence, deep surveys with SKA-1 has the potential to detect substantial numbers of such bubbles up to fairly highly redshifts, although we stress that the exact numbers would depend on the details of the reionization scenario.

In Figure~\ref{bubble_sizes}, we plot the number of ionizing photons $N_\mathrm{ion,tot}$ that have gone into various ionized bubbles of volume $V_\mathrm{ion}$ at $z=7$--$11$ in our simulations. Due to large density fluctuations on small scales, there is substantial variation (by more than one order of magnitude) in the number of ionizing photons that have been used to produce the smaller bubbles ($V_\mathrm{ion}\sim 10^1$--$10^2$ cMpc$^3$). However, as bubbles approach the SKA-1 resolution limit ($V_\mathrm{ion}\gtrsim 10^3$ cMpc$^3$), the effects of density fluctuations tend to even out, leaving a $1\sigma$ range that corresponds to a factor of $<4$ in the required $N_\mathrm{ion,tot}$ for a fixed comoving $V_\mathrm{ion}$. This suggests that SKA-1 measurements of the volumes of ionized bubbles may relatively tight limit on the number of ionizing photons that have been emitted into the IGM within these regions, thereby allowing for constraints on the properties of the galaxy populations within these structures. Sect~\ref{sim_params} features a brief discussion on how these results are affected by different assumptions on the ionization efficiency. 

The best-fitting $N_\mathrm{ion,tot}-V_\mathrm{ion}$ relation varies slightly between the different redshift snapshots, but combining the simulation data from all snapshots with significant numbers of $V_\mathrm{ion}\gtrsim 1000$ cMpc$^3$ bubbles gives the average relation:
\begin{equation}
N_\mathrm{ion,tot}\approx 8\times 10^{67} \left(\frac{V_\mathrm{ion}}{\mathrm{cMpc}^3}\right) ^{1.03}.
\label{Nion_Vion_relation}
\end{equation}
We will adopt this relation in the following sections to predict the number of galaxies required to produce a bubble of a given volume. 

Note that in an exact inside-out reionization scenario, analytically one would expect the power index in Eq.~\eqref{Nion_Vion_relation} to be $\sim 1$, which is consistent with the results from our simulations. The scatter in the power law index from panel to panel of Fig.~\ref{bubble_sizes} is mainly due to the spatial fluctuations in the hydrogen number density, clustering of the sources and non-conservation of the ionizing photon numbers in the later part of the reionization \citep{choudhury18}.

\section{How the ionizing photon budget within IGM bubbles is tied to properties of bubble galaxies}
\label{galaxies}
The results presented in section~\ref{sizes} suggest that the volumes of the isolated bubbles that SKA-1 will be able to resolve are strongly coupled to the number of ionizing photons that have been emitted throughout the previous history of these regions. In section~\ref{sim_params}, we also argue that this number is relatively insensitive to how the production of ionizing photons is distributed across the halo population. 

In our fiducial simulations, ionized bubbles of the smallest size that SKA-1 can hope to resolve ($V_\mathrm{ion}\sim 10^3$ cMpc$^3$) include $\sim 1000$ dark matter halos of mass $\gtrsim 10^9\ M_\odot$, which represents a reasonable ballpark estimate of the {\it total} number of galaxies expected within these structures (but please note that only a very small fraction of these will be sufficiently bright to be detected). However, the exact number of halos or galaxies needed to produce the required number of ionizing photons will depend on how efficient these are in emitting ionizing photons into the IGM. If the galaxies produce very few ionizing photons (e.g. because of intermittent star formation) or if only a small fraction of the ionizing photons enter the IGM (due to low $\langle f_\mathrm{esc}\rangle$), then only very extreme matter overdensities, with more halos and more galaxies, will be able to produce resolvable bubbles. Vice versa, if galaxies are highly efficient in emitting ionizing photons into the IGM, then resolvable bubbles will contain fewer halos and galaxies. 

We note that, under the assumption of an invariant stellar initial mass function, the $N_\mathrm{ion,tot}$ parameter is closely tied to the total mass $M_\mathrm{stars}$ locked up in stars. For the set of simulated galaxies and the spectral evolutionary model adopted in this paper (see Sect.~\ref{Nion_LUV_section}), the approximate relation is:
\begin{equation}
M_\mathrm{stars} \approx 1\times10^{11}\left( \frac{N_{\rm ion,tot}}{1\times 10^{71}}\right) \left(\frac{\langle f_\mathrm{esc} \rangle}{0.1}\right)^{-1} \ M_\odot
.    
\label{Mstars_Nion_eq}
\end{equation}

In principle, this stellar mass could be locked up within a single galaxy, but this would require a very extreme scenario. If we assume $\langle f_\mathrm{esc} \rangle \leq 0.1$ and consider a galaxy that starts forming stars somewhere in the  $z\approx 15$--20 range and manages to do so at a constant star formation rate thereafter, then the $\approx 1\times 10^{71}$ ionizing photons required to produce a $V_\mathrm{ion}\approx 10^3$ cMpc$^3$ bubble (eq.~\ref{Nion_Vion_relation}) by $z\approx 10$ would correspond to a total stellar mass of $\geq 1\times 10^{11} \ M_\odot$, a star formation rate (SFR) $\geq 500 M_\odot$ yr$^{-1}$ and a dust-free UV luminosity $M_\mathrm{UV}\lesssim -25.0$. Such bright, high-mass galaxies are not yet known at $z\gtrsim 8$, and to bring such objects in agreement with the brightest galaxies known in this redshift range \citep{Calvi16,Stefanon19} would required $>2$ mag of UV dust attenuation. In this section, we will therefore assume that the ionized IGM bubbles that SKA-1 can resolve contain a population of galaxies, rather than a single object that somehow formed in isolation, and proceed to discuss the details of how the ionizing photon budget of a bubble translates into estimates of the number of detectable galaxies within this structure.

\subsection{Past production of ionizing photons tied to the rest-frame UV luminosity}
\label{Nion_LUV_section}
\begin{figure*}
\includegraphics[scale=0.4]{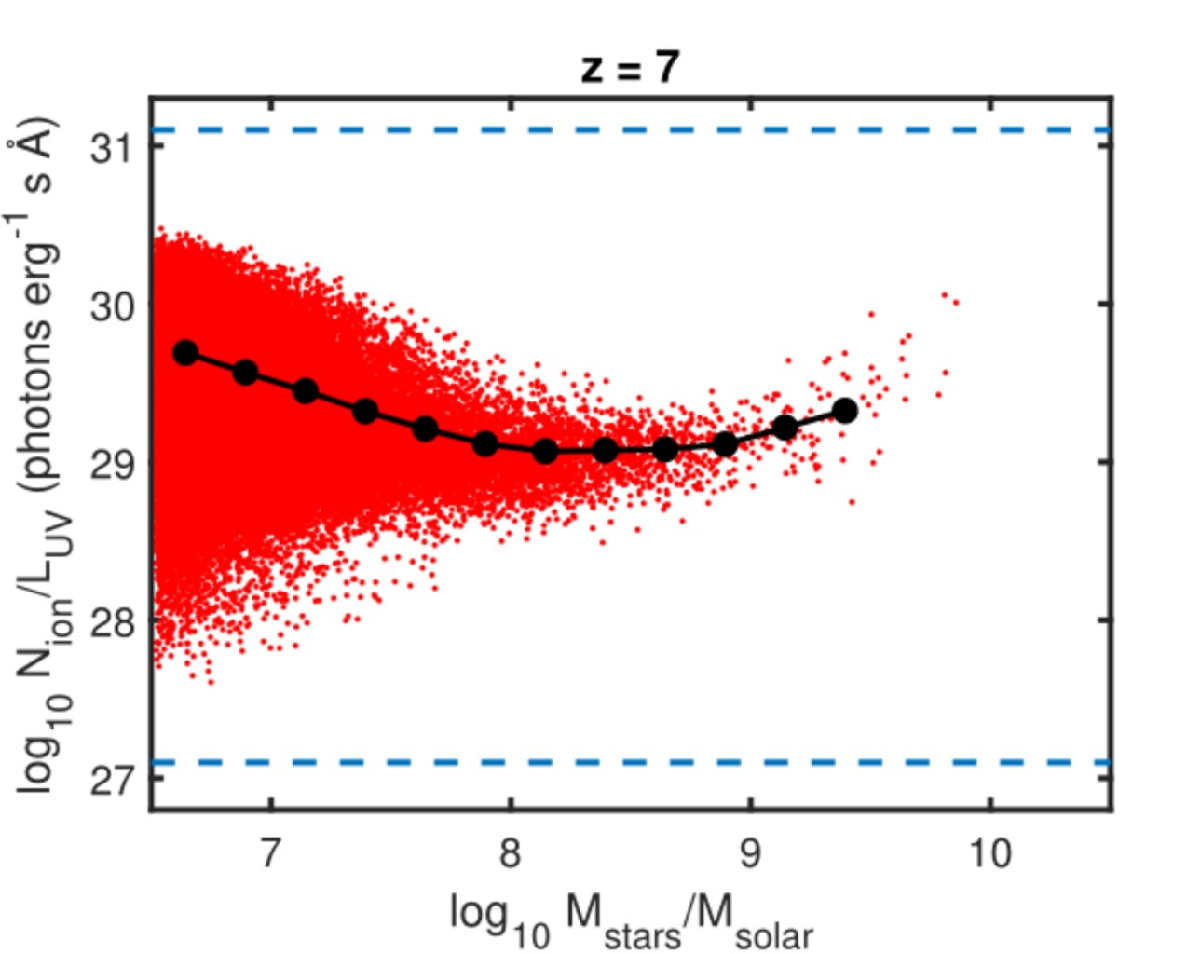}
\includegraphics[scale=0.395]{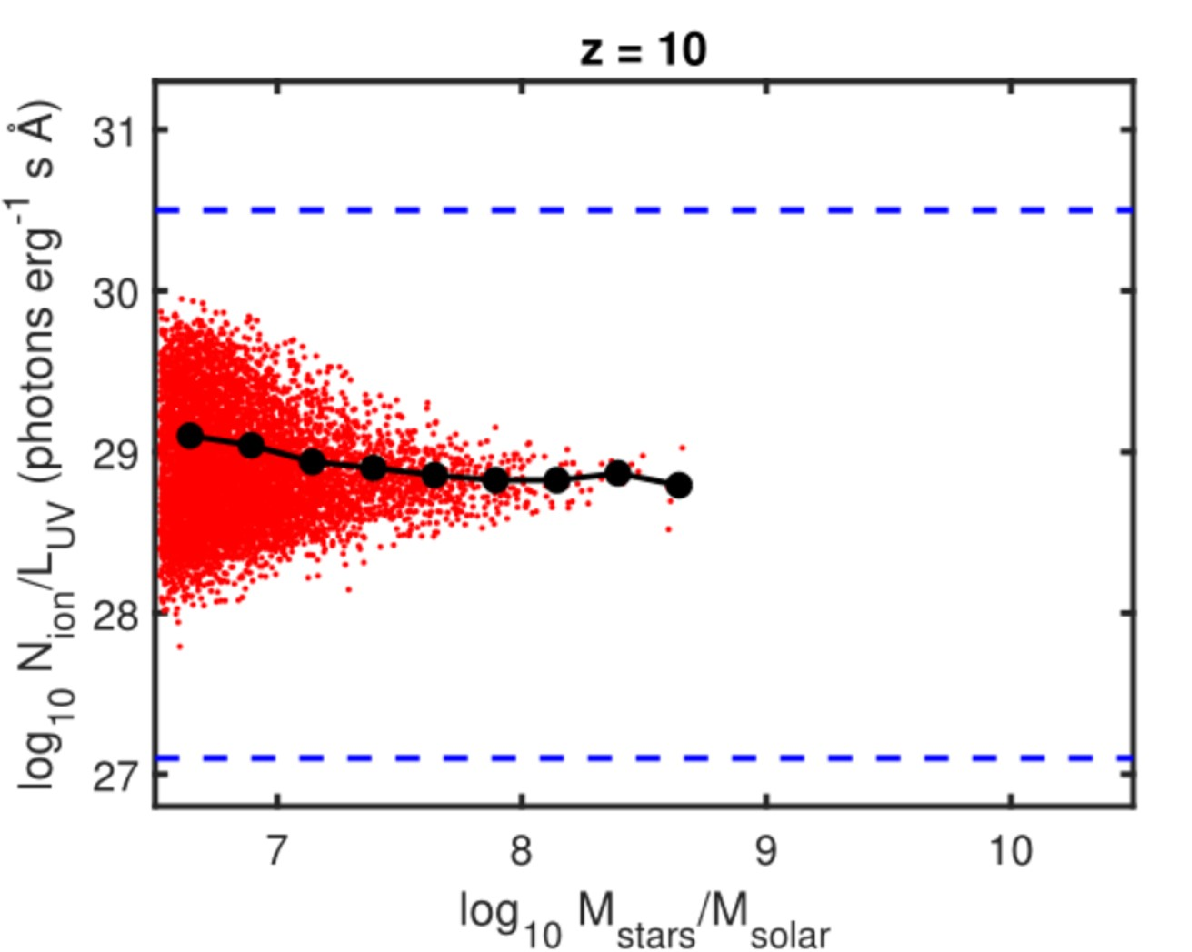}
\caption{Ratio between the cumulative number of ionizing photons produced by a galaxy and its momentary rest-frame UV 1500~\AA{} luminosity, as a function of its total stellar mass at $z=7$ (left) and $z=10$ (right). Red dots represent galaxies from the \citet{Shimizu16} simulations. The solid lines with filled circles indicate how the arithmetic mean evolves with galaxy mass and the dashed horizontal lines represent the minimum and maximum $N_\mathrm{ion}/L_\mathrm{UV}$ ratios theoretically allowed at this redshift.}
\label{Nion_Mstars}
\end{figure*}

By aiming a telescope with near-IR capabilities (e.g. Euclid, JWST, WFIRST, ELT) at the same area of the sky surveyed by SKA-1 for 21 cm emission at $z>6$, we can detect the rest-frame UV\footnote{In the case of JWST, also the rest-frame optical will be within reach.} ($\lambda \gtrsim 1216$ \AA{}) light from galaxies in these structures. Throughout this paper, we will quantify the UV luminosity $L_\mathrm{UV}$ of $z>6$ galaxies using the monochromatic luminosity or flux at a rest-frame of 1500 \AA{}.  The UV luminosity measured this way reflects the recent star formation rate over the past $\sim 10$--$100$ Myr \citep[e.g.][]{Boquien14}.  For star formation histories stretching over several billions years, as in the case of low-redshift galaxies, this would not be a good proxy for the total stellar mass or the total number of ionizing photons ever produced by this object, since the prior star formation rate could have either been much higher or much lower than in the epoch from which we detect its light.

However, simulations of reionization-epoch galaxies generically predict that $z>6$ galaxies should experience semi-continuous star formation, often with star formation rates increase over time for the more massive ones \citep[e.g.][] {Finlator11,Jaacks12,Dayal13,Shimizu14,Ma15}. Semi-continuous star formation, coupled to the limited time span since the onset of star formation (a few hundred Myr) in the $z>6$ galaxy population, limits the variations one can expect in the ratio between $N_\mathrm{ion}$, the cumulative number of ionizing photons a galaxy has produced in the past (either in situ or within smaller galaxies that have merged into this galaxy by the redshift at which it is observed), and $L_\mathrm{UV}$. 
Low-mass galaxies may well experience more stochastic star formation activity \citep[e.g.][]{Mutch16,Ma18}, and consequently larger variations between $N_\mathrm{ion}$ and $L_\mathrm{UV}$, but the greater number density of such objects also means that such variations may largely average out over a bubble population that contains large numbers of galaxies. For the interested reader, appendix~\ref{Appendix_A} features a more thorough description of how this $N_\mathrm{ion}/L_\mathrm{UV}$ parameter is tied to the prior star formation history.

By combining the \citet{Shimizu16} simulations for $z=7$ and $z=10$ galaxies with the stellar population spectra produced with the Starburst99 model \citep{Leitherer99} under the assumption of the \citet{Kroupa01} universal IMF and Geneva stellar evolutionary tracks with high mass-loss, \citet{Calzetti00} dust attenuation and nebular emission as in \citet{Zackrisson17}, we predict the distribution of $N_\mathrm{ion}/L_\mathrm{UV}$ ratios as a function of total stellar mass $M_\mathrm{stars}\geq 10^{6.5}\ M_\odot$ in  Figure~\ref{Nion_Mstars} for $z=7$ and $z=10$. For models with a standard stellar initial mass function, the LyC escape fraction $f_\mathrm{esc}$ has no significant impact on the 1500 \AA{} luminosity and here it has been set to $f_\mathrm{esc}=0$. 

As seen in Figure~\ref{Nion_Mstars}, the $N_\mathrm{ion}/L_\mathrm{UV}$ ratio displays galaxy-to-galaxy variations by factors of a few at the highest masses, but varies by more than two orders of magnitude among the lowest-mass galaxies resolved ($\log_{10} (M_\mathrm{stars}/M_\odot)\approx 6.5$) due to large temporal fluctuations in star formation activity within these objects. Such low-mass galaxies are expected to contribute most to the ionizing photon budget within an ionized IGM bubble, but are also present in larger numbers, which means that summing the fluctuating $N_\mathrm{ion,i}$ contributions for the whole population of bubble galaxies still results in a fairly well-constrained $N_\mathrm{ion,tot}$. 

The mean $N_\mathrm{ion}/L_\mathrm{UV}$ ratio (solid line) also evolves slightly with $M_\mathrm{stars}$ and reaches its highest value for the smallest $M_\mathrm{stars}$ due to the increasingly stochastic star formation rates of such objects. High $N_\mathrm{ion}/L_\mathrm{UV}$ ratios are produced by galaxies which have experienced a high SFR in the past, but are observed in a phase when the star formation activity is very low -- leading to a near-constant $N_\mathrm{ion}$ set by the prior activity and a fading $L_\mathrm{UV}$ due to the aging stellar population.

To put these $N_\mathrm{ion}/L_\mathrm{UV}$ ratios into context, a very bright $M_\mathrm{UV} \approx -20$ ($L_\mathrm{UV} \approx 6\times 10^{40}$ erg s$^{-1}$ \AA{}$^{-1}$) galaxy at $z=10$ with $N_\mathrm{ion}/L_\mathrm{UV}\approx 6\times 10^{28}$ photons erg$^{-1}$ s \AA{} would have produced $4\times 10^{69}$ ionizing photons over its lifetime, which -- by itself -- is insufficient (by more than an order of magnitude) to produce an ionized bubble that SKA-1 can resolve (requires $\sim 10^{71}$ ionizing photons) even in the case of $\langle f_\mathrm{esc} \rangle\approx 1$. For $\langle f_\mathrm{esc} \rangle \approx 0.1$, it would take $\approx 300$ such galaxies to produce a detectable bubble. However, given the shape of the halo mass function or the galaxy luminosity function, it is far more likely that an even larger number of much fainter galaxies is present within these structures.

While it is possible that $N_\mathrm{ion}/L_\mathrm{UV}$ ratios even larger than seen in Figure~\ref{Nion_Mstars} may be relevant for galaxies below the resolution limit of the simulation used, the $N_\mathrm{ion}/L_\mathrm{UV}$ ratio cannot fluctuate without bounds. In the absence of stellar IMF variations, the lower limit would be set by a newborn stellar population (age $\approx 1$ Myr) which for a Starburst99, $Z=0.004$ stellar population of the type adopted here is $\log_{10 }N_\mathrm{ion}/L_\mathrm{UV} \approx 27.1$ photons erg$^{-1}$ s \AA{}, whereas the upper limit would be set by an instantaneous-burst population (a.k.a. a single or simple stellar population) with an age equal to the age of the Universe. At $z=7$ and $z=10$, this limit would be at $\log_{10 }N_\mathrm{ion}/L_\mathrm{UV} \approx 30.5$ and 31.1 photons erg$^{-1}$ s \AA{} respectively. These theoretical limits are indicated by dashed lines in Figure~\ref{Nion_Mstars}.

In the following, however, we will assume that the $N_\mathrm{ion}/L_\mathrm{UV}$ ratio follows a base-10 lognormal function with $\mu\approx 29.34$ (28.89),  $\sigma\approx 0.46$ (0.34) and arithmetic means $N_\mathrm{ion}/L_\mathrm{UV} \approx 3.7\times 10^{29}$ ($1.1\times 10^{29}$) photons erg$^{-1}$ s \AA{}  at $z=7$ ($z=10$). While this approach fails to capture the evolution of the mean $N_\mathrm{ion}/L_\mathrm{UV}$ ratio with mass evident from Figure~\ref{Nion_Mstars}, this does not have any substantial impact on the distribution unless assumptions on $\langle f_\mathrm{esc} \rangle$ places very large weight on galaxies in some particular mass range (see Section~\ref{discussion}). At $z=7$, the mean $N_\mathrm{ion}/L_\mathrm{UV}$ ratio varies by a factor of $\approx 4$ across the range of galaxy masses considered, and limiting the contribution to the bubble ionization to some very narrow current mass range could in principle alter the result by  up to this factor. At $z=10$, the corresponding factor is $\approx 2$. Given the large total number of galaxies ($\gtrsim$ 1000; see Section~\ref{total_number_of_galaxies}) that in our model are expected to inhabit the ionized IGM bubbles that SKA-1 can detect, the impact of the mass evolution of the object-to-object scatter in  $N_\mathrm{ion}/L_\mathrm{UV}$ around the mean is negligible compared to the mass evolution of the mean $N_\mathrm{ion}/L_\mathrm{UV}$ itself.

\subsection{Total number of galaxies per bubble}
\label{total_number_of_galaxies}
To predict galaxy number counts within individual IGM bubbles, we will adopt the simplifying assumption that galaxies within an ionized IGM bubble exhibit higher number densities but are otherwise similar to field galaxies at the same redshift (see Sect.~\ref{assembly_bias} for a discussion on this). We adopt the relative shape of the $z \approx 7$ and $z\approx 10$ UV luminosity functions by \citet{Bouwens15}, extended down to $M_\mathrm{UV}=-14$,  and randomly sample the scatter predicted in the case of the \citet{Shimizu16} simulation in Figure~\ref{Nion_Mstars}. We then calculate the number of galaxies necessary to produce the $N_\mathrm{ion,tot}\approx 1\times 10^{71}$ photons (eq.~\ref{Nion_Vion_relation}) required to obtain a $V_\mathrm{ion}\approx 10^3$ cMpc$^3$ ionized bubble by $z\approx 7$ and $z\approx 10$. Following this procedure we obtain $N_\mathrm{galaxies} \langle f_\mathrm{esc} \rangle \approx 260 $ galaxies at $z\approx 7$ and $\approx 1300$ at $z\approx 10$. Hence, for $\langle f_\mathrm{esc} \rangle \approx 0.1$, we would expect a total of $\approx 2600$ galaxies in a bubble resolvable by SKA-1 at $z\approx 7$, and $\approx 13000$ galaxies at $z\approx 10$.  The value is higher at $z\approx 10$ due to a combination of lower $N_\mathrm{ion}/L_\mathrm{UV}$ and differences in the luminosity function. The scatter in $N_\mathrm{ion}/L_\mathrm{UV}$  just affects these estimate at the $\approx 10\%$ level compared to adopting a constant $N_\mathrm{ion}/L_\mathrm{UV}$ throughout the whole galaxy population. 

While we have here adopted $M_\mathrm{UV}=-14$ as the faint cut-off of the $z=7$--10 luminosity function, observations of lensed fields have indicated that it may in fact extend several magntiudes fainter than this before turning over \citep[e.g.][]{Bouwens16,Livermore17}. The effect of adopting a fainter cut-off limit would would boost the total number of bubble galaxies, thus further reducing the effects of galaxy-to-galaxy scatter in $N_\mathrm{ion}/L_\mathrm{UV}$. For instance, assuming that bubble galaxies are forming down to $M_\mathrm{UV}=-10$ (while keeping the same luminosity function shape) would boost the {\it total} number of galaxies by a factor of $\approx 30$, but has a much smaller effect on the number of detectable galaxies, as will be demonstrated in the next section. The factor of $\approx 30$ is smaller than would be expected from a simple extrapolation of the luminosity function to fainter magnitudes, since this extension alters the ionizing photon flux budget and requires a different absolute scaling of the luminosity function.

\subsection{Galaxy detection limits}
\label{detection_limits}
Only a small fraction of the galaxies present within an ionized IGM bubble ($\lesssim1\%$ by number) are likely to appear above the detection threshold of near-IR telescopes within the foreseeable future. 

To provide quantitative estimates for the number of detectable galaxies, we consider both photometric detections with Euclid, WFIRST and JWST plus spectroscopic detections with ELT/MOSAIC and JWST/NIRSpec. The pros and cons of these two detection methods are described in more detail in Section~\ref{spec_vs_phot}, but the basic difference is that spectroscopic surveys are less prone to line-of-sight interlopers, whereas photometric surveys in principle can probe further down the galaxy luminosity function within the bubble. Below, we describe the various detection limits we consider in the discussion on detectability of bubble galaxies. When assessing the detection of emission lines, we have chosen to be conservative and therefore ignore the Ly$\alpha$ line. Even though the ionized IGM in SKA-selected bubbles may well allow a favourable transmission factor of Ly$\alpha$ photons through the IGM, scattering and extinction within the galaxies may still render this line very weak for many of these objects. 

{\bf Euclid}\footnote{\url{http://sci.esa.int/euclid/}} is a 1.2 m telescope scheduled for launch in 2022 with optical and near-IR imaging capabilities that can also do 1.1-2.0 micron slitless spectroscopy (resolution $\lambda/\Delta(\lambda)=250$). While Euclid will provide a survey of 15,000 deg$^2$, Euclid deep fields of about 40 deg$^2$ degrees in total will also be observed, with $5\sigma$ broadband detection limits in the optical of $m_\mathrm{AB}\approx 27$ mag and $m_\mathrm{AB}\approx 26$ in the $YJH$ bands. Galaxy candidates at $z>6$ can be singled out through drop-out criteria in multiband surveys of this type, by requiring these candidates to be undetected in all filters that sample their spectra at wavelengths shortward of the redshifted Ly$\alpha$ break, yet detected in one or several filters on the longward side of the break. Throughout this paper, we will assume that a sufficient dropout criterion is met if an object is undetected at the $2\sigma$ level shortward of the Ly$\alpha$ break, yet detected at $5\sigma$ or more in at least one filter on the other side. For the multiband imaging surveys considered in this paper, we neglect any minor variation in flux detection thresholds among the different near-IR bands, and therefore simply adopt the $5\sigma$ limit as the effective dropout detection threshold ($m_\mathrm{AB}\approx 26$ in the case of Euclid). 
The line detection limit for Euclid is estimated at $\approx 5\times 10^{-17}$ erg cm$^{-2}$ s$^{-1}$ \citep{Marchetti17}. For galaxies at $z\gtrsim 7$, Euclid can cover lines in the rest-frame UV up to $\lambda \leq 2500$ \AA{}, which basically covers HeII (1640 \AA{}), CIV (1549 \AA{}), OIII] (1666 \AA{}), CIII] (1909 \AA{}). However, these lines are in general expected to remain undetectable at $m_{AB}\approx 26$ mag for a $\approx 5\times 10^{-17}$ erg cm$^{-2}$ s$^{-1}$ spectroscopic detection limit \citep[e.g.][]{Shimizu16}, which means that Euclid will only detect $z\gtrsim 7$ objects as faint as $m_{AB}\approx 26$ mag through imaging. Hence, we only consider a $m_{AB}\approx 26$ mag photometry threshold of this telescope.

{\bf WFIRST}\footnote{\url{https://www.nasa.gov/wfirst}} is a 2.4 m telescope scheduled for launch in the mid-2020s, which is envisioned to be equipped with imaging capabilities in the 0.48--2.0 micron range and a slitless spectroscopy mode covering 1.00--1.95 micron. WFIRST will carry out wide-field surveys (2200 deg$^2$ in the high-latitude survey), but also deep field ($\approx 20$ deg$^2$) observations with expected imaging and spectroscopy detection limits of $m_\mathrm{AB}\approx 28$ mag and $\approx 1\times 10^{-17}$ erg cm$^{-2}$ s$^{-1}$ respectively. Here, we consider only the imaging detection limit, since the line flux detection limit will effectively be much brighter than the $m_\mathrm{AB}\approx 28$ mag limit \citep[as predicted by the simulations of][]{Shimizu16}. 

{\bf James Webb Space Telescope}\footnote{\url{https://jwst.nasa.gov/}} (JWST), scheduled for launch in 2021, is a 6.5 m telescope that will be able to do extremely deep imaging and spectroscopy in the 0.6-5 micron range and will hence have access to the rest-frame optical lines from $z\approx 7$--13 galaxies that neither Euclid nor WFIRST will. The downside is the much smaller field of view, which is 3.6\arcmin $\times$ 3.4\arcmin{} for the JWST/NIRSpec spectrograph and 4.4\arcmin $\times$ 2.2\arcmin{} for JWST/NIRCam (imaging or spectroscopy), which implies $\approx 2$--3 fields to cover just the smallest ionized bubbles that SKA-1 can resolve ($\approx 5$\arcmin across). If we consider a total of $\approx 20$ ($\approx 100$) hours of exposure time to cover a single ionized bubble with either photometry or spectroscopy, we arrive at a point-source detection limit of $m_\mathrm{AB}\approx 29$ (30) mag for NIRCam imaging in two short-wavelength channel (0.6--2.2 $\mu$m) filters across two fields (although photometry in two long-wavelength channel filters at 2.5--5 $\mu$m would also be achieved simultaneously). The corresponding limits for spectroscopy are redshift-dependent, because the most suitable emission line ([OIII] (5007 \AA{}) at $z=7$ is out of JWST/NIRSpec range at $z=10$. Using the \citet{Shimizu16} model to predict the emission line strengths for galaxies with a given UV contiuum flux, we instead base the $z=10$ limit on the [OII] (3727 \AA{}) line at $z=10$. For observing programmes of either $\approx 20$ hours or $\approx 100$ hours across two fields in resolution $R\approx 1000$ mode, this results in S/N$\approx$ 5 line detection limits of $3\times 10^{-19}$ erg s$^{-1}$ cm$^{-2}$ or $1.3\times 10^{-19}$ erg s$^{-1}$ cm$^{-2}$, which corresponds to galaxies of $m_\mathrm{AB}\approx 28.3$ or $\approx 29.2$ mag at $z=7$, but $m_\mathrm{AB}\approx 26.5$ or $m_\mathrm{AB}\approx 27.5$ at $z=10$. The continuum detection limits are approximately the same if the CIII] (1909 \AA{}) line is targeted instead of [OII] at $z=10$.

{\bf Extremely Large Telescope}\footnote{\url{https://www.eso.org/sci/facilities/eelt/}} (ELT) is the largest groundbased optical/near-IR telescope under construction and will have first light around 2025. The currently planned ELT instrumentation does not allow for wide-field imaging, so we do not consider this option in the present paper. However, the planned MOSAIC instrument, which is expected to be operational towards the end of the 2020s, is expected to be capable of multi-object spectroscopy at 0.9--1.8 micron over a 7 arcmin diameter field. Hence, MOSAIC can cover the smallest ionized bubbles detected by SKA-1 in just one fields and should, in a total of 40 hours of observing time, be able to detect rest-frame UV lines at $z\approx 7$ with S/N=5 at $\approx 1\times 10^{-19}$ erg cm$^{-2}$ \citep{Evans15}. By targeting the CIII] (1909 \AA{}) line at $z=7$ and the CIV (1549 \AA{}) at $z=10$, this corresponds to galaxies with $m_\mathrm{AB}\approx 29.0$ mag at $z=7$ and $\approx 28.25$ mag at $z=10$, based on predictions from the \citet{Shimizu16} models. However, we stress that pre-imaging at this depth will be required to select the spectroscopic targets for ELT/MOSAIC, and this imaging cannot easily be performed by ELT itself given currently planned instrumentation.

\subsection{Detectable galaxies per bubble}
\label{detectable}
\begin{figure*}
\includegraphics[width=\columnwidth]{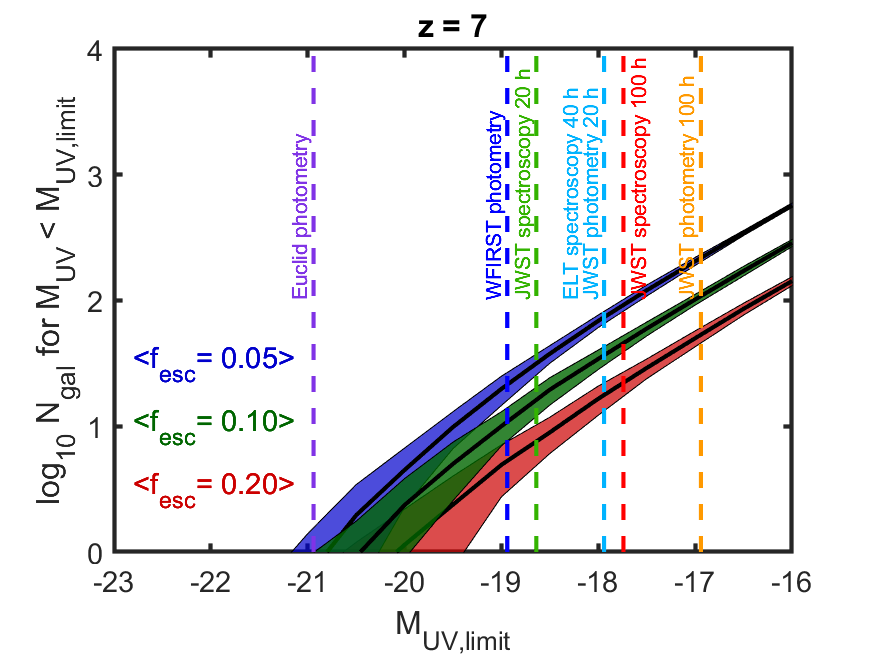}
\includegraphics[width=\columnwidth]{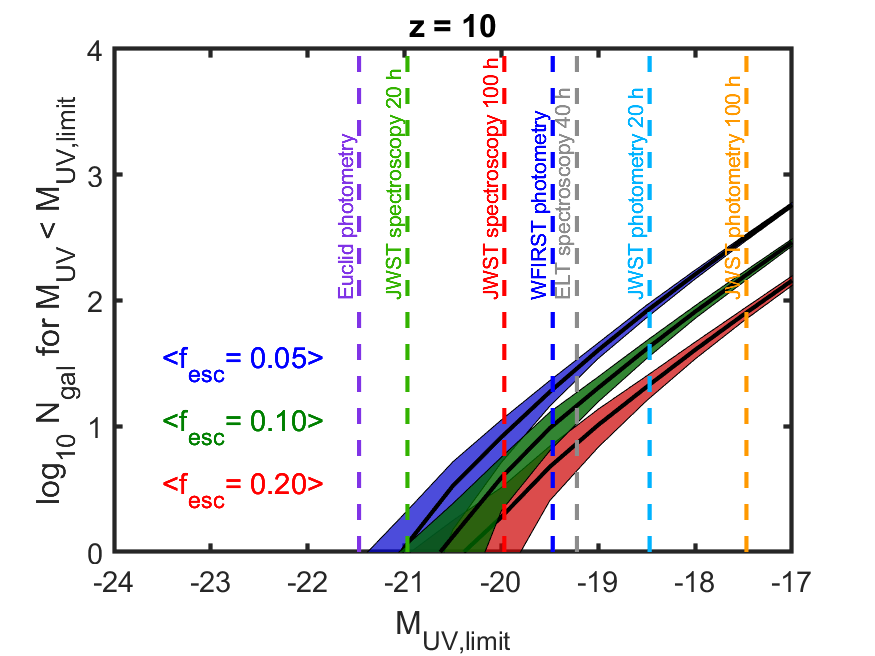}
\caption{Number of galaxies expected above different limiting rest-frame UV absolute magnitude within a $V_\mathrm{ion}\approx 1000$ cMpc$^3$ ionized bubble (set by the requirement that $\approx 1\times 10^{71}$ ionizing photons need to be emitted into the IGM) at $z=7$ (left) and $z=10$ (right), for $\langle f_\mathrm{esc} \rangle=0.05$ (blue stripes), 0.1 (green stripes) or 0.2 (red  stripes). The width of the stripes is set by the predicted standard deviation in the galaxy number counts between individual bubbles, caused by the random sampling of the adopted luminosity function with faint cut-off at $M_\mathrm{UV}=-14$. Even in the most pessimistic case ($f_\mathrm{esc}=0.2$, one expects to detect tens of galaxies above the deepest detection limits. However, even the most optimistic predictions ($\langle f_\mathrm{esc} \rangle=0.05$) indicate that Euclid could well be blind to galaxies in the smallest ionized IGM bubbles that SKA-1 may resolve.  
\label{Ngal_fesc}}
\end{figure*}

In Figure~\ref{Ngal_fesc} we show the number of galaxies expected above these various detection limits for thre different options concerning the time-integrated, photon-number weighted LyC escape fraction $\langle f_\mathrm{esc} \rangle$. Here, we have assumed that the bubble galaxies follow a luminosity function with the same relative shape (but different scaling) as the \citet{Bouwens15} $z \approx 7$ and $z\approx10$ UV luminosity function, extended down to either $M_\mathrm{UV}=-14$ or $M_\mathrm{UV}=-10$. Under the assumption that $\langle f_\mathrm{esc} \rangle$ has no mass/luminosity dependence (see Section~\ref{fesc} for a discussion on this), the number of galaxies expected above the various detection threshold is given by the differently stripes, for $\langle f_\mathrm{esc} \rangle=0.05$, 0.1 and 0.2. 

How would these results change if we assume that the luminosity function retains its shape faintward of $M_\mathrm{UV}=-14$?
Our computational machinery indicates that, at fixed $\langle f_\mathrm{esc} \rangle$, one expects to detect a factor of $\approx 2$ fewer galaxies at $z=7$ (a factor of $\approx 3$ at $z=10$) within an ionized bubble if the luminosity function is extended down to $M_\mathrm{UV}=-10$. The conversion can simply be done by shifting all the plotted galaxy counts down by this factor. It should however be noted that this parametrization assumes that galaxies all the way down to $M_\mathrm{UV}=-10$ display $N_\mathrm{ion,tot}/L_\mathrm{UV}$ ratios that follow a lognormal distribution with parameters similar to those presented in Section~\ref{Nion_LUV_section}. However, galaxies as faint as $M_\mathrm{UV}=-10$ may have total stellar masses as low as $M_\mathrm{stars}\sim 10^5 \ M_\odot$, which is significantly below the resolution limit of our simulations. The intermittent star formation episodes expected in such low-mass systems may well cause a shift in the mode of the distribution, which can be explored with higher-resolution simulations in the future.

There are a few things to note from Figure~\ref{Ngal_fesc}. For reasonable values of $\langle f_\mathrm{esc}\rangle$ ($\approx 0.05$--0.2), considerable numbers of potentially detectable galaxies are expected within each bubble at both $z=7$ and $z=10$, and this number scales with $1/\langle f_\mathrm{esc} \rangle$, since a higher $\langle f_\mathrm{esc} \rangle$ means that fewer galaxies are required to provide the ionizing photons needed to form the bubble. Even in the most pessimistic case shown ($\langle f_\mathrm{esc} \rangle = 0.2$), one expects to detect a handful of galaxies at $z=7$ with WFIRST photometry and several tens of galaxies with either JWST photometry programme. Spectroscopy with ELT or JWST can also produce $\sim 10$ bubble galaxy detections. 

At $z=10$, JWST spectroscopy fares somewhat worse than ELT spectroscopy because the intrinsically brighter [OIII] line that gave JWST an edge at $z=7$ have been redshifted out of JWST range. The pessimistic limits at $z=10$ places several tens of galaxies above the detection limit of the JWST imaging surveys, and a few galaxies above the threshold of either WFIRST imaging or ELT spectroscopy.

The detection prospects for Euclid are considerably worse, and we conclude that Euclid may largely be blind to dropout galaxies in the smallest ionized IGM bubbles that SKA-1 can resolve. 

The difference between the $z=7$ and $z=10$ cases is that, due to the lower $N_\mathrm{ion}/L_\mathrm{UV}$ at $z=10$ (see Section~\ref{Nion_LUV_section}) and a slightly different shape of the adopted luminosity function, a larger number of galaxies is required at $z=10$ than at $z=7$ to produce a bubble of a given size, provided that $\langle f_\mathrm{esc} \rangle$ is kept fixed. The detection limits are also slightly shifted to brighter UV luminosities at $z=10$.

Because $\langle f_\mathrm{esc} \rangle$ is here assumed to be independent of UV luminosity (see Section~\ref{discussion} for a discussion on this), only a minor fraction (e.g. $\approx$ 7--20\% for WFIRST, but up to $\approx$ 20--50\%  for the deepest JWST photometry limits) of the ionizing photons that have contributed to the ionization of the IGM in the bubble are accounted for by galaxies above the detection limits in the case where the luminosity function is truncated at $M_\mathrm{UV}=-14$.

\begin{figure}
\includegraphics[width=\columnwidth]{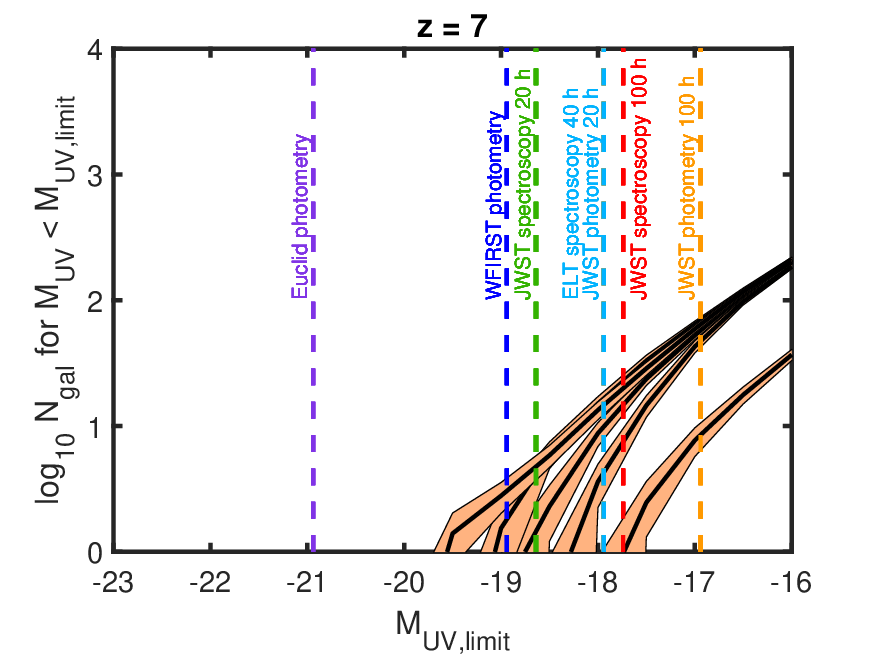}
\caption{Same as the left panel of Fig.~\ref{Ngal_fesc}, but with predictions based on halo catalogs from five different realizations of $\approx 1000$ cMpc$^3$ bubbles drawn from the semi-numerical simulations of Section~\ref{bubble_section}, coupled to an analytical recipe for coupling halo mass to UV luminosity. Each orange stripe corresponds to the predicted galaxy counts for one bubble, with the stripe width representing the standard deviation in galaxy number counts stemming from the scatter in the relation between halo mass and UV luminosity. This procedure gives rise to a luminosity function for the bubble galaxies that differs in shape from the one assumed in  Fig.~\ref{Ngal_fesc}, with fewer galaxies that would be detectable above the brighter detection limits but mostly a similar number at the faintest detection limit as the predictions for $\langle f_\mathrm{esc} \rangle \approx 0.2$ in Fig.~\ref{Ngal_fesc}. However, one of the bubbles has managed to reach $\approx 1000$ cMpc$^3$ volume with a significantly lower number of total ionizing photons than the rest, implying much lower galaxy counts. \label{Ngal_fesc_halocat}}
\end{figure}

As an alternative to the procedure used to generate the galaxy count predictions of Fig.~\ref{Ngal_fesc}, in which the relative shape of the observationally determined field luminosity functions was used as a basis for populating IGM bubbles with galaxies until a fixed ionizing photon budget had been reached, one may instead start from the halo mass distribution. In Fig.~\ref{Ngal_fesc_halocat}, we start from the dark matter halos predicted within the $\approx 1000$ cMpc$^3$ bubbles predicted by the fiducial reionization simulations of Section~\ref{bubble_section} at $z=7$ and then adopt the fitting function presented by \citet{Inoue18} for the relation (and Gaussian scatter) between halo mass and UV continuum luminosity to attach galaxy fluxes to each halo. If we furthermore adopt the $N_\mathrm{ion}/L_\mathrm{UV}$ distribution of Section~\ref{Nion_LUV_section}, we would arrive at $\langle f_\mathrm{esc} \rangle \approx 0.15$ for these galaxies. However, the predicted number of bright galaxies ($M_\mathrm{UV}<-19$) in Fig.~\ref{Ngal_fesc_halocat} is several times lower than would be expected from even $\langle f_\mathrm{esc} \rangle \approx 0.2$, in Fig.~\ref{Ngal_fesc}, whereas the number of galaxies at the faintest detection levels ($M_\mathrm{UV}\approx -17$) in most cases are similar to what one estimate for $\langle f_\mathrm{esc} \rangle \approx 0.15$ through interpolation in Fig.~\ref{Ngal_fesc_halocat}. This is primarily because
the procedure of relating UV luminosities to halo masses in Fig.~\ref{Ngal_fesc_halocat} results in a luminosity function for bubble galaxies with a different shape than that assumed in Fig.~\ref{Ngal_fesc}, with more very faint galaxies at the expense of bright ones. Moreover, this procedure also gives rise to substantial bubble-to-bubble variations in predicted galaxy counts, due to differences in halo mass distributions within bubbles of similar volume. To illustrate this, Fig.~\ref{Ngal_fesc_halocat} features galaxy predictions for five randomly selected $\approx 1000$ cMpc$^3$ bubbles. One of these bubbles stands out in having reached this volu  me despite a much smaller combined dark halo mass and ionzing photon production than the rest, resulting in a galaxy population that would only be detectable at the very faintest detection limits considered. 

As further discussed in Section~\ref{assembly_bias}, more sophisticated simulations, meeting observational constraints on the properties of $z\gtrsim 7$ galaxies in overdense regions, will be require to gauge which of the two approaches (Figure ~\ref{Ngal_fesc} or \ref{Ngal_fesc_halocat}) that results in the most realistic estimates.

\subsection{Photometric or spectroscopic selection?}
\label{spec_vs_phot}

Bubble galaxies may be identified either in an imaging/photometry survey or through spectroscopy. For a given telescope and a fixed total observing time, photometry will typically reach deeper, but drop-out criteria (or photometric redshifts based on an SED fit from multiband data) have the drawback of not allowing very accurate redshift information. For a typical broadband drop-out criterion, the redshift error will be $\Delta(z)\approx 1$. This should be compared to the size of ionized IGM bubbles, which for a spherical 1000 cMpc$^3$ bubble, will cover a line-of-sight depth that at $z=6$--10 will be $\Delta(z)\approx 0.03$--0.06. Hence, an imaging survey runs the risk of misidentifying galaxies located in the foreground or background as bubble members. A spectroscopic survey, on the other hand, would only need a relatively low spectral resolution of $R=\lambda/\Delta(\lambda)\gtrsim 200$ to reach the redshift accuracy required to identify the bubble membership of a given galaxy through the detection of an identified emission line.

\begin{figure}
\includegraphics[width=\columnwidth]{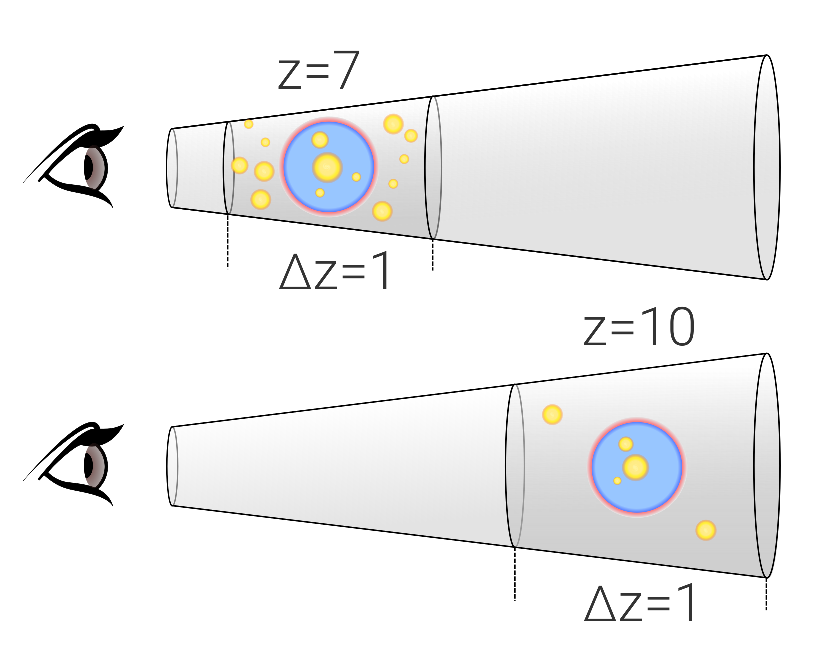}
\caption{Schematic figure illustrating the issue of interlopers in a drop-out survey (here assumed sensitive to galaxies within a redshift interval of $\Delta(z)=1$) towards an ionized bubble at either redshift $z=7$ or $z=10$. The number of galaxies inside an ionized bubble of fixed size detectable by an imaging survey are expected to change just by a factors of a few between $z\approx 7$ and $z\approx 10$, whereas the number of interlopers in line-of-sight $\Delta(z)=1$ volume the is expected be significantly higher at lower redshift. Hence, a photometric survey becomes less risky when aimed at a bubble at higher redshift (here $z\approx 10$) then when aimed at a bubble at the end of reionization ($z\approx 7$).
\label{schematic_los}}
\end{figure}

How substantial is the risk of misidentifications in a photometric survey? This depends on the redshift of the bubble targeted. From the detection limits in Fig.~\ref{Ngal_fesc}, we see that the number of detectable galaxies above a certain luminosity limit, within a bubble of fixed size, changes by no more than a factor of $\approx 3$ between $z\approx 7$ and $z\approx 10$ if a constant $\langle f_\mathrm{esc} \rangle$ is assumed. At the same time, the ambient number density of galaxies above a certain threshold luminosity drops by an order of magnitude between these redshifts \citep{Bouwens15}. This leads to a situation (schematically illustrated in Figure~\ref{schematic_los}) where the number of interlopers in a drop-out survey towards a given bubble will be much greater towards the end of reionization ($z\approx 7$) than at earlier stages ($z\approx 10$).

Another way to understand this is to note that ionizing a $\sim 1000$ cMpc$^3$ volume requires a similar ionizing photon budget, and hence a similar collapsed mass, at both redshifts (Fig.~\ref{bubble_sizes}). However, a much higher peak in the density field is needed for such a structure to collapse and form stars by $z\approx 10$ than at $z\approx 7$. Such high peaks are correspondingly rarer, and their overdensity compared to their environment is much greater. For a bubble with a line-of-sight depth of $\Delta(z)\approx 0.05$, the volume probed by a broadband imaging survey (line-of-sight resolution $\Delta(z)\approx 1$) at $z=7$--10 will be more than 20 times larger than that of the bubble, but the average galaxy number density in this volume is also likely to be much lower, since the overdense regions tend to be the first to reionize. If we adopt the cosmic average for the number density of galaxies in the line-of-sight volume outside the bubble, the \citet{Bouwens15} luminosity function predicts that there should at $z\approx 7$ be $\approx 25$ interloper galaxies at $M_\mathrm{UV}\leq -19.5$  in the $\Delta(z)\approx 1$ cylindrical volume projected against a 1000 cMpc$^3$ bubble. This is larger than the number of bubble galaxies for all the $\langle f_\mathrm{esc} \rangle$ cases considered at $z=7$ in Fig.~\ref{Ngal_fesc}. However, at $z=10$, the number of interloper galaxies at $M_\mathrm{UV}\leq -19.5$ (approximately the WFIRST photometry limit at this redshift) is $\approx 1$, which means that photometric interlopers within the $\Delta(z)=1.0$ volume selected by drop-out criteria are unlikely to be a problem, since the number of bubble galaxies predicted in Fig.~\ref{Ngal_fesc} is expected to be several times higher than this.

A consistency check of this conclusion, based on halo statistics instead of luminosity functions, is presented in Appendix~\ref{Appendix_B}.

We caution, however, that a purely photometric survey may also be prone to interlopers from much lower redshifts than indicated by the $\Delta(z)=1$ volume considered above, because very strong optical emission lines can give the appearance of a Ly$\alpha$ break, unless the selection is based on detection on several filters shortward of the break. 

\section{Discussion}
\label{discussion}

\subsection{Galaxy assembly bias in overdense regions}
\label{assembly_bias}
Throughout this paper, we have -- as a first baseline approach -- adopted the assumptions that the relative shape of the galaxy luminosity function at $z\geq 7$ is independent of environment, that the ionized bubbles of the IGM simply feature a scaled-up version of the field luminosity function at the same redshift, and that there are no significant differences between the properties of galaxies residing in a matter overdensity and those in the field. However, overdense regions that reionize early are, by definition, not typical and may well have galaxy populations quite different from the cosmic average at this epoch.

At low redshifts, it is well established that the luminosity function and the ratio of red- to blue-sequence galaxies change with environment \citep[e.g.][]{McNaught-Roberts14}. Changes in galaxy properties with environment have observationally been traced up to $z\approx 3$ \citep[e.g.][]{Grutzbauch11}, but exactly how early such differences get imprinted in the galaxy population remains an open question \citep[for a review, see][]{Overzier16}.

In simulations, both the shape of the dark halo mass function and the properties of individual halos of a given mass (in terms of accretion rate, spin, concentration and shape) are predicted to be affected by the overdensity of the environment \citep{Lee17}, and increased merger rates, galaxy interactions and the feedback from ionizing radiation produced within overdense regions may further augment changes in galaxy properties compared to the field population. 

A common expectation is that feedback from an ultraviolet background may quench star formation in low-mass dark matter halos \citep[e.g.][]{Mesinger08,Sobacchi13,Maio16}. In an ionized bubble, this could potentially alter the shape of the galaxy luminosity function at the faint end, or alternatively affect the typical ionizing emissivity of faint galaxies. To first order, this can be treated as an effective truncation of the luminosity function (below which galaxies do not contribute ionizing photons) similar to the different faint luminosity function extensions that we have considered in this paper ($M_\mathrm{UV}$ limit -14 to -10). However, if external feedback also affects the star-forming properties of massive galaxies \citep[for a scenario of this type, see][]{Susa08}, then changes to the shape of the luminosity function at the bright end may also occur. Assembly bias (the notion that the statistical properties of galaxies depend on properties other than halo mass) could also manifest itself in other, more complex ways. Indeed, some simulations have indicated differences in specific star formation rates, galaxy mass functions, metallicities and dust content in overdense regions compared to field at $z\geq 6$ \citep{Yajima15,Sadoun16}. 

As samples of $z>7$ galaxies grow larger, it should be possible to observationally test for such environmental effects by, for instance, studying the slope of the bright end of the luminosity function as a function of clustering. A couple of notable overdensities of galaxies at $z\gtrsim 7$ have already been discovered in deep HST surveys -- a $z\approx 7$ overdensity of 17 Lyman-break galaxy candidates, out of which three are confirmed Lyman-alpha emitters (two with consistent redshifts) in an area a few arcminutes across \citep{Castellano16,Castellano18}, and a $z\approx 8.4$ overdensity of up to 8 Lyman-break galaxy candidates, out of which one is a confirmed Lyman-alpha emitter, in an area just a ten arcseconds across \citep{Ishigaki16,Laporte17}. Provided that a fair fraction of these Lyman-break galaxies are at the redshift of Lyman-alpha emitters, the brightness distribution of Lyman-break galaxies in these regions is roughly consistent with what our models predict for ionized regions of a scale that SKA-1 should be able to resolve. Recently, an overdensity of 12 Lyman-$\alpha$ emitters at $z\approx 6.6$ was discovered by \citet{Harikane19}, but these appear to cover a volume that is considerably larger than covered by our simulations. By targeting such overdensities with deep photometric and spectroscopic JWST observations it may be possible to observationally constrain feedback and environmental effects within some of the most extreme overdensities in the reionization epoch. A few years down the line, many more such overdensities are also expected in to be uncovered in WFIRST deep field observations, which are expected to cover $\sim 100$ times the solid angle of correspondingly deep HST surveys. 

In future efforts, it would be worthwhile to study the properties of simulated galaxies drawn directly from ionized IGM regions of reionization simulations to study the effects that assembly bias and galaxy feedback are expected to have on galaxy counts within such structures. Recently, \citet{Geil17} used such simulations to study how the sizes of ionized bubbles correlate with the luminosity of the brightest galaxy within the bubble. Their study focuses on ionized IGM bubbles at $z\approx $9--11 that are factors of a few smaller in radius than the SKA-1 tomographic resolution limit than we have adopted, but if we apply their best-fitting relation between bubble radius and brightest bubble galaxy at $z\approx 10$ to bubbles of radius 2--3 cMpc (similar to their mean bubble size at this redshift) to derive the UV luminosity of their brightest galaxy in these structures, and then scale the number of such galaxies ($M_\mathrm{UV}=-18.3$ to -19.5) up by factors of $\approx 9$--30 to match the volume of our $\approx 1000$ cMpc$^3$ bubbles, we get a good match with our $\langle f_\mathrm{esc} \rangle=0.2$ predictions for this luminosity range at $z=10$ in Figure~\ref{Ngal_fesc}. While this indicates that there is agreement for the {\it average} bubble, the extreme tail of their distribution contains galaxies that are much brighter than any of our predictions in Figure~\ref{Ngal_fesc} -- for example, their very largest bubble (radius $\approx 4.8$ cMpc) at $z\approx 11$ contains a brightest galaxy that attains $M_\mathrm{UV}=-22.5$ (albeit under the assumption of zero dust attenuation) which seems to completely dominate the UV luminosity of its bubble (as seen in their Figure A1), violate our assumption of a relatively well-sampled bubble luminosity function and lie closer to the monolithical galaxy case discussed in Section~\ref{galaxies}. Whether even rarer and more extreme objects could exist that produce individual ionized IGM bubbles sufficiently large to be detectable with SKA-1 warrants further study.

\subsection{Population III stars and active galactic nuclei}
Our current treatment assumes that ionizing photons from young stars represent the only substantial contribution to the ionizing photon budget within 21 cm bubbles. However, if objects or mechanisms other than star-forming galaxies and their associated Lyman continuum radiation contribute to bubble growth, this could of course affect the outcome. 

Population III stars are expected to emit a large fraction of their radiation at Lyman continuum energies, owing both to high stellar surface temperatures at zero metallicity and potentially a top-heavy stellar initial mass function \citep[e.g.][]{Schaerer02}. However, Population III stars in minihalos are only likely to dominate the cosmic star formation rate density at $z\gtrsim 15$ \citep[e.g.][]{Maio10} and are not expected to play any major, global role during late stages of cosmic reionization \citep[e.g.][]{Kulkarni14}. Even so, one could envision such stars to have a significant effect {\it locally}, in regions of delayed and highly concentrated Population III star formation. 

It has been argued that Population III galaxies may form in $z<15$ HI cooling halos that have managed to remain chemically pristine \citep{Stiavelli10}, but current simulations suggest that such systems are unlikely to attain total stellar masses higher than $\sim 10^6\ M_\odot$ \citep[e.g.][]{Yajima17b,Inayoshi18}. Based on the Population III galaxy SED models of \citet{Zackrisson11}, such systems cannot, over the expected lifetimes of $\sim 10^6$--$10^8$ yr, produce more than $\sim 10^{70}$ ionizing photons even under the assumption of an extremely top-heavy stellar initial mass function (typical stellar mass $\sim 100\ M_\odot$). Since $\sim 10^{71}$ photons are required to produce a bubble detectable with SKA-1, such rare, exotic galaxies cannot contribute significantly to the photon budget, even if they exhibited extreme Lyman continuum leakage ($f_\mathrm{esc}\approx 1$). 

This leaves black hole accretion as the most likely mechanism to rival the ionizing flux from star-forming galaxies within bubbles close to the SKA-1 resolution limit. A single quasar can easily produce $\sim 10^{71}$ ionizing photons in as little as $\sim 10^7$ yr \citep[e.g.][]{Maselli07}. An active, high-luminosity quasar  would usually be readily identifiable as such from spectroscopy, but an accreting black hole that has contributed early on and then has turned dormant may not be. Is it then possible that SKA-1 may detect ionized IGM bubbles that appear completely devoid of galaxies when probed with upcoming near-IR telescope, because the quasar responsisible for the bubble is in an inactive phase? This could potentially happen for the most shallow limits considered in Figure~\ref{Ngal_fesc} but seems unlikely for the deepest ones. Using scaling relations from \citet{Wood&Loeb}, a supermassive black hole radiating at the Eddington luminosity for $\sim 10^7$ yr with $f_\mathrm{esc}=1.0$ would need to have a mass of $\sim 10^8\ M_\odot$ to seriously impact the ionizing photon budget of a 1000 cMpc$^3$ bubble. For a bulge-to-supermassive black hole mass ratio of $\lesssim 0.1$ at $z\leq 10$ \citep{Targett12}, this requires a host galaxy with a stellar mass of at least $\gtrsim 10^9\ M_\odot$, which is expected to have a luminosity $M_\mathrm{UV}<-19$ and therefore be in the detectable range of many of the bubble surveys considered in Figure~\ref{Ngal_fesc}. Moreover, statistics on galaxy counts within a few bubbles of similar size can put constraints on scenarios including such transient, stochastic contributions to bubble growth; and the sharpness and morphology of the 21 cm profile of the bubble may also reveal contributions from X-ray photons that could be inconsistent with a normal stellar population \citep[e.g.][]{Tozzi00,Wyithe07,Pacucci14,Ghara16,Kakiichi17}. 

\subsection{The ionization efficiency of halos}
\label{sim_params}
The fiducial reionization simulations used to produce $N_\mathrm{ion,tot}$-$V_\mathrm{ion}$ predictions of Fig.~\ref{bubble_sizes} are based on the assumption of a fixed ionization efficiency $n_{\rm ion}$ for all halos above a minimum halo mass of $M_{{\rm halo,\, min}} = 1.09 \times 10^9\, {\rm M_{\odot}}$. Raising (lowering) the adopted ionization efficency alters the reionization history by shifting the completion of reinization to higher (lower) redshifts  \citep[see Fig. 1 of ][]{Greig17b}, but this effect can be offset by simultaneously raising (lowering)  $M_{{\rm halo,\, min}}$. Through this degeneracy, current observational constraints on cosmic reionization are consistent with  ionization efficencies that differ by an order of magnitude \citep{Greig17a,Monsalve18}. 

One could perhaps suspect that our results concerning the connection between the volume ($V_\mathrm{ion}$) of ionized bubbles and the number of ionizing photons produced therein ($N_\mathrm{ion,tot}$) would strongly depend on the overall scaling of $n_{\rm ion}$ or its halo mass dependence. However, while changing the relation between $n_{\rm ion}$ and $M_{\rm halo}$ does affect the reionization history of the simulations, we find that the bubble population remains strongly correlated with the overall ionization state of the universe, (i.e. the global neutral fraction ($\bar{x}_{{\rm HI}}$) quoted in different panels of Figure~\ref{bubble_sizes} for our standard parameter set), and that both the average $N_\mathrm{ion,tot}$-$V_\mathrm{ion}$ relation and the scatter around this relation remains almost the same for $V_\mathrm{ion}\geq 1000$ cMpc$^3$ despite rather substantial alterations of the relation between $n_{\rm ion}$ and $M_{\rm halo}$. Specifically, by redistributing the production of ionizing photons across the halo population, while still retaining the same reionization history ($\bar{x}_{{\rm HI}}(z)$) as in Figure~\ref{bubble_sizes}, we find that the scatter at  $V_\mathrm{ion}\geq 1000$ cMpc$^3$ is insignificantly altered when either the minimum halo mass for ionizing photon production is raised to $M_{{\rm halo,\, min}} = 10^{10}\, {\rm M_{\odot}}$, when a random uniform scatter that corresponds to a factor of 20 variation in ionizing efficiency is added to each halo, or when the relation between the ionizing photons produced by a halo($\propto M_\mathrm{halo}^n$) is altered from $n=1$ to $n=1.41$.

\subsection{The impact of small-scale density variations}
\label{clumping}
A shortcoming in the semi-numerical simulations that we have used to obtain the relation between $V_\mathrm{ion}$ and $N_\mathrm{ion,tot}$ in Section \ref{sizes} is that this machinery does not take small-scale spatial variations of hydrogen density into account when calculating the recombination rate. Instead, the recombination rate is assumed to be uniform throughout the IGM (independent of density). This ignores the effect of the clumping factor and hence the relatively rapid recombinations that may take place in high-density, non-star forming regions known as damped Lyman-$\alpha$ absorbers (DLA) or Lyman limit systems (LLS). These DLAs and LLSs are expected to be dense enough to self-shield themselves from the ionizing background created by the galaxies and quasars and work as the sinks of ionizing radiation within an ionized bubble. Under the standard model of structure formation in our universe, these systems are expected to be more abundant than the halos that are capable of hosting galaxies that we assume produces majority of the ionizing photons. A failure to properly consider such small-scale, high-density systems therefore likely leads us to underestimate the $N_\mathrm{ion,tot}$ required to produce bubbles of a given volume $V_\mathrm{ion}$. 

To accurately take into account the effects of these absorbers when predicting the ionization topology and the corresponding ionizing photon budget would require reioniziation simulations spanning a very large dynamic range. On one hand they should have mass resolutions of the order of Jeans scales to model the recombination processes inside these sub-Mpc objects and on the other hand they should be able to simulate large volumes (of the order of Gpc) to be able produce the bubble size distribution and the corresponding large scale fluctuations in the signal. Recent results from theoretical modeling \citep{schaye01, kaurov15}, high resolution but sub-Mpc size simulations \citep{park16} and their sub-grid adaptation in the large scale semi-numerical simulations \citep{Sobacchi14} suggest that non-uniform recombinations in the IGM would cost $2$-$3$ or more photons per ionized hydrogen atom by the end of the reionization era. A direct and obvious implication of this on the  $N_\mathrm{ion,tot}$ vs $V_\mathrm{ion}$ plot shown in Figure \ref{bubble_sizes}, will be an increment in the amplitude of the power law fit at all stages of reionization. This would  also increase the scatter in the plot at the low $V_\mathrm{ion}$ end of the plots. However, as we approach $V_\mathrm{ion} \sim 1000\,{\rm Mpc}^3$ (our adopted SKA-1 detection limit), one would expect this scatter to die down substantially due to the effect of averaging over large volumes. This should be explored more carefully in future simulations.

\subsection{Multiple ionizations from each ionizing photon}
Our current treatment assumes that each hydrogen-ionizing (Lyman continuum)  photon emitted from stars will ionize exactly one hydrogen atom. However, ionized gas may itself emit ionizing photons through free-bound and free-free transitions, effectively resulting in multiple ionizations from a single stellar Lyman continuum photon. The overall impact of this depends on the shape of the ionizing stellar continuum and the gas temperature, but is not expected to boost the effective ionizing emissivity of galaxies by more than at most a factor of 1.6 under realistic conditions \citep[][]{Inoue10}. 

\subsection{The escape fraction of ionizing photons}
\label{fesc}
Throughout this paper, we have assumed the same $\langle f_\mathrm{esc} \rangle$ for all bubble galaxies, but one can also envision this parameter evolving as a function of UV luminosity, halo mass or stellar mass \citep[e.g][]{Yajima11,Kimm&Cen14,Paardekooper15,Xu16,Sharma17}. This will affect the number of galaxies required to emit a fixed number of ionizing photons into the IGM, but the detectable number of bubble galaxies may still be tied to the time-integrated, photon-number weighted $\langle f_\mathrm{esc} \rangle$ of the whole bubble population. This population-wide $\langle f_\mathrm{esc} \rangle$ quantity basically represents the ratio of the total number of ionizing photons that have ever escaped into the IGM over the total number produced within the bubble, and the results of Figure~\ref{Ngal_fesc} still hold as long as $\langle f_\mathrm{esc} \rangle$ is interpreted this way. For example, in the case where the $z=7$ luminosity function is considered to extend to $M_\mathrm{UV}=-14$, a scenario in which the very faintest galaxies with $M_\mathrm{UV}>-16$ have $\langle f_\mathrm{esc} \rangle=0.25$ and the rest have $f_\mathrm{esc}=0$ produces a population-wide $\langle f_\mathrm{esc} \rangle \approx 0.1$ and the same number of detectable galaxies as predicted by $\langle f_\mathrm{esc} \rangle \approx 0.1$ in that figure. Similarly, a scenario with the same luminosity function in which bright galaxies with $M_\mathrm{UV}\leq -19$ have $\langle f_\mathrm{esc} \rangle=0.25$ and fainter galaxies have $\langle f_\mathrm{esc} \rangle=0.0$ corresponds to a population-wide $\langle f_\mathrm{esc} \rangle \approx 0.05$ and the same number of detectable galaxies as indicated by that line. 

In principle, it would seem that using SKA-1 to estimate the volume of an ionized bubble and then simply counting the galaxies brighter than a specific $M_\mathrm{UV}$ detection threshold within that structure would result in an observational constraint on the $\langle f_\mathrm{esc} \rangle$ parameter. However, this would only be possible insofar as both the shape of bubble galaxy luminosity function above and below the $M_\mathrm{UV}$ detection limit and the $N_\mathrm{ion}/L_\mathrm{UV}$ parameter are under control, i.e. that assembly/environmental biases of the type discussed in Section~\ref{assembly_bias} are either negligible or can be accurately modelled. Moreover, the measurement error on the SKA-1 volume would also need to be $\lesssim 50\%$ to make this measurement meaningful -- a very difficult endeavour indeed. 

\subsection{Studying larger bubbles}
In the current paper, we have focused on ionized bubbles of size $\approx 1000$ cMpc$^3$, which is at the limit of what SKA-1 can hope to detect. In future works, it would in future works also be interesting to consider larger bubbles, since these would be easier to detect in the 21 cm data due to higher signal-to-noise. Providing a census on the galaxy content of such structures with near-IR telescopes would on the other hand be more challenging, since the observing time required to survey the galaxy content of such bubbles to the same depth scales with the projected area of the bubble in the plane of the sky. Hence, a spherical bubble that has a volume a factor of 5 times larger requires a factor of $5^{(2/3)}\approx 3$ more observing time if the same magnitude limit is to be reached. On the other hand, larger structures are also likely to contain brighter galaxies, which would make the most extreme members detectable with telescopes wide-field survey telescopes like Euclid. 

\section{Conclusions}
\label{summary}
Our results can be summarized as follows:
\begin{itemize}
\item Our semi-numerical simulations indicate that -- for ionized IGM bubbles sufficiently large to be resolved by  SKA-1 (volume $\gtrsim 1000$ cMpc$^3$) -- there is a relatively tight relation between the volume of ionized IGM bubbles at $z=7$--10 and the number of ionizing photons that have escaped from the galaxies within. For bubbles of volume $\gtrsim 1000$ cMpc$^3$, the $1\sigma$ bubble-to-bubble scatter in this relation is a factor of $< 4$ (Section~\ref{sizes}). 

\item The total number of ionizing photons required to produce an ionized IGM bubble sufficiently large to be resolved by SKA-1 (volume $\gtrsim 1000$ cMpc$^3$) at $z>6$ is $\gtrsim 1\times 10^{71}$. This can be converted into a rough constraint on the minimum total stellar mass that has formed within such a structure (eq.~\ref{Mstars_Nion_eq}), and -- using additional assumptions on the properties of the bubble galaxies --- estimates on the number of galaxies detectable within that structure at the redshift where the bubble is detected (Section~\ref{galaxies}). 

\item Using conservative assumptions on the shape of the luminosity function of bubble galaxies, the prior star formation history of these objects and their time-integrated, photon number-weighted mean Lyman continuum escape fractions, we predict that the deepest spectroscopic surveys with JWST or ELT of SKA-detected ionized bubbles at $z=7$--10 are expected to turn up at least a few bubble galaxies (and in many cases far more), even in the case of the smallest resolvable bubbles. The same also holds for purely photometric (multiband-imaging) observations with JWST or WFIRST. However, Euclid may only be able to detect galaxies within ionized bubble with volumes an order of magnitude higher than the SKA-1 detection limit (Section~\ref{detectable}). Detailed detection predictions are presented in Figure~\ref{Ngal_fesc} and Figure~\ref{Ngal_fesc_halocat}.

\item Whereas spectroscopic observations (with JWST or ELT) may be required to survey the smallest SKA-detected bubbles towards the end of reionization ($z\approx 7$) without ending up with excessive numbers of line-of-sight interlopers, photometric surveys (with WFIRST or JWST) may be competitive at higher redshifts ($z\approx 10$) due to the smaller number of field-galaxy  interlopers expected at redshifts close to that of the bubble (Section~\ref{spec_vs_phot}).

\item If large numbers of bubble galaxies are detected, this could in principle allow for combined constraints on the luminosity function of bubble galaxies and the time-integrated, photon number-weighted mean escape fraction $\langle f_\mathrm{esc} \rangle$ of ionizing photons from these objects into the surrounding IGM (Section~\ref{fesc}). For instance, if one adopts our baseline assumption -- which seems consistent with the current observational picture of the $z\gtrsim 7$ galaxy population -- that galaxies in ionized IGM bubbles are not significantly different (in terms of star formation history and relative shape of the luminosity function) from field galaxies at the same redshift, it should be possible to derive a lower limit on $\langle f_\mathrm{esc} \rangle$ by adopting a very faint extension of the luminosity function when assessing the galaxy contribution to the ionizing photon budget. However, effects related to galaxy assembly bias and environmental feedback in overdensities could complicate this procedure  (Section~\ref{assembly_bias}). Such effects may manifest themselves as differences between bubble and field galaxies in terms of spectral properties, luminosity function shape and in the ratio between luminous and dark halo mass. 

The detection of such signatures, which would shed new light on the properties of galaxies that form in extreme environments of the $z\gtrsim 7$ Universe, may be uncovered by through the bubble galaxy observations themselves (e.g. spectral and dynamical properties uncovered through JWST and ELT spectroscopy; luminosity function constraints through WFIRST or JWST photometry). Early indications (prior to the detection of any ionized IGM bubbles with SKA-1) of such effects at $z\gtrsim 7$ of may also come from studies with ALMA, JWST and WFIRST of currently known overdensities of $z\gtrsim 7$ galaxies, since some of these structures are likely located in ionized IGM bubbles that SKA-1 will eventually be able to resolve though 21 cm tomography. The theoretical understanding of such differences between galaxies in overdensites and the field, and how they affect the emergence of ionized IGM bubbles, must ultimately come from numerical simulations geared to this specific problem.

\end{itemize}

\section*{Acknowledgements}
EZ acknowledges funding from the Swedish National Space Agency. EZ, SM, AD and GM would like to acknowledge financial assistance through the SPARC scheme (sponsored by MHRD, India) under the project titled ``Imaging the first billion years of the universe with next-generation telescope". RM would like to acknowledge funding form the Science and Technology Facilities Council [grant numbers ST/F002858/1 and ST/I000976/1] and the Southeast Physics Network~(SEPNet). MS acknowledges funding from Stiftelsen Olle Engkvist Byggm\"{a}stare. PD acknowledges support from the European Research Council's starting grant ERC StG-717001 (``DELPHI") and from the European Commission's and University of Groningen's CO-FUND Rosalind Franklin program. UM was supported through a research grant awarded by the German Research Fundation (DFG), project n. 390015701. AM acknowledges support by the European Research Council (ERC) under the European Union's Horizon 2020 research and innovation programme (grant agreement No 638809 -- AIDA). The results presented here reflect the authors' views; the ERC is not responsible for their use.




\bibliographystyle{mnras}
\bibliography{refs} 

\begin{thebibliography}{}
\makeatletter
\relax
\def\mn@urlcharsother{\let\do\@makeother \do\$\do\&\do\#\do\^\do\_\do\%\do\~}
\def\mn@doi{\begingroup\mn@urlcharsother \@ifnextchar [ {\mn@doi@}
  {\mn@doi@[]}}
\def\mn@doi@[#1]#2{\def\@tempa{#1}\ifx\@tempa\@empty \href
  {http://dx.doi.org/#2} {doi:#2}\else \href {http://dx.doi.org/#2} {#1}\fi
  \endgroup}
\def\mn@eprint#1#2{\mn@eprint@#1:#2::\@nil}
\def\mn@eprint@arXiv#1{\href {http://arxiv.org/abs/#1} {{\tt arXiv:#1}}}
\def\mn@eprint@dblp#1{\href {http://dblp.uni-trier.de/rec/bibtex/#1.xml}
  {dblp:#1}}
\def\mn@eprint@#1:#2:#3:#4\@nil{\def\@tempa {#1}\def\@tempb {#2}\def\@tempc
  {#3}\ifx \@tempc \@empty \let \@tempc \@tempb \let \@tempb \@tempa \fi \ifx
  \@tempb \@empty \def\@tempb {arXiv}\fi \@ifundefined
  {mn@eprint@\@tempb}{\@tempb:\@tempc}{\expandafter \expandafter \csname
  mn@eprint@\@tempb\endcsname \expandafter{\@tempc}}}

\bibitem[\protect\citeauthoryear{{Bagley} et~al.,}{{Bagley}
  et~al.}{2017}]{Bagley17}
{Bagley} M.~B.,  et~al., 2017, preprint, \href
  {http://adsabs.harvard.edu/abs/2017arXiv170105193B} {} (\mn@eprint {arXiv}
  {1701.05193})

\bibitem[\protect\citeauthoryear{{Barkana}}{{Barkana}}{2016}]{Barkana16}
{Barkana} R.,  2016, \mn@doi [\physrep] {10.1016/j.physrep.2016.06.006}, \href
  {http://adsabs.harvard.edu/abs/2016PhR...645....1B} {645, 1}

\bibitem[\protect\citeauthoryear{{Beardsley}, {Morales}, {Lidz}, {Malloy}  \&
  {Sutter}}{{Beardsley} et~al.}{2015}]{Beardsley15}
{Beardsley} A.~P.,  {Morales} M.~F.,  {Lidz} A.,  {Malloy} M.,   {Sutter}
  P.~M.,  2015, \mn@doi [\apj] {10.1088/0004-637X/800/2/128}, \href
  {http://adsabs.harvard.edu/abs/2015ApJ...800..128B} {800, 128}

\bibitem[\protect\citeauthoryear{{Boquien}, {Buat}  \& {Perret}}{{Boquien}
  et~al.}{2014}]{Boquien14}
{Boquien} M.,  {Buat} V.,   {Perret} V.,  2014, \mn@doi [\aap]
  {10.1051/0004-6361/201424441}, \href
  {http://adsabs.harvard.edu/abs/2014A%26A...571A..72B} {571, A72}

\bibitem[\protect\citeauthoryear{{Bouwens} et~al.,}{{Bouwens}
  et~al.}{2015}]{Bouwens15}
{Bouwens} R.~J.,  et~al., 2015, \mn@doi [\apj] {10.1088/0004-637X/803/1/34},
  \href {http://adsabs.harvard.edu/abs/2015ApJ...803...34B} {803, 34}

\bibitem[\protect\citeauthoryear{{Bouwens}, {Oesch}, {Illingworth}, {Ellis}  \&
  {Stefanon}}{{Bouwens} et~al.}{2016}]{Bouwens16}
{Bouwens} R.~J.,  {Oesch} P.~A.,  {Illingworth} G.~D.,  {Ellis} R.~S.,
  {Stefanon} M.,  2016, preprint, \href
  {http://adsabs.harvard.edu/abs/2016arXiv161000283B} {} (\mn@eprint {arXiv}
  {1610.00283})

\bibitem[\protect\citeauthoryear{{Calvi} et~al.,}{{Calvi}
  et~al.}{2016}]{Calvi16}
{Calvi} V.,  et~al., 2016, \mn@doi [\apj] {10.3847/0004-637X/817/2/120}, \href
  {http://adsabs.harvard.edu/abs/2016ApJ...817..120C} {817, 120}

\bibitem[\protect\citeauthoryear{{Calzetti}, {Armus}, {Bohlin}, {Kinney},
  {Koornneef}  \& {Storchi-Bergmann}}{{Calzetti} et~al.}{2000}]{Calzetti00}
{Calzetti} D.,  {Armus} L.,  {Bohlin} R.~C.,  {Kinney} A.~L.,  {Koornneef} J.,
   {Storchi-Bergmann} T.,  2000, \mn@doi [\apj] {10.1086/308692}, \href
  {http://adsabs.harvard.edu/abs/2000ApJ...533..682C} {533, 682}

\bibitem[\protect\citeauthoryear{{Castellano} et~al.,}{{Castellano}
  et~al.}{2016}]{Castellano16}
{Castellano} M.,  et~al., 2016, \mn@doi [\apjl] {10.3847/2041-8205/818/1/L3},
  \href {http://esoads.eso.org/abs/2016ApJ...818L...3C} {818, L3}

\bibitem[\protect\citeauthoryear{{Castellano} et~al.,}{{Castellano}
  et~al.}{2018}]{Castellano18}
{Castellano} M.,  et~al., 2018, \mn@doi [\apjl] {10.3847/2041-8213/aad59b},
  \href {http://esoads.eso.org/abs/2018ApJ...863L...3C} {863, L3}

\bibitem[\protect\citeauthoryear{{Choudhury} \& {Paranjape}}{{Choudhury} \&
  {Paranjape}}{2018}]{choudhury18}
{Choudhury} T.~R.,  {Paranjape} A.,  2018, \mn@doi [\mnras]
  {10.1093/mnras/sty2551}, \href
  {https://ui.adsabs.harvard.edu/\#abs/2018MNRAS.481.3821C} {481, 3821}

\bibitem[\protect\citeauthoryear{{Choudhury}, {Haehnelt}  \&
  {Regan}}{{Choudhury} et~al.}{2009}]{choudhury09}
{Choudhury} T.~R.,  {Haehnelt} M.~G.,   {Regan} J.,  2009, \mn@doi [\mnras]
  {10.1111/j.1365-2966.2008.14383.x}, \href
  {http://adsabs.harvard.edu/abs/2009MNRAS.394..960C} {394, 960}

\bibitem[\protect\citeauthoryear{{Datta}, {Bharadwaj}  \& {Choudhury}}{{Datta}
  et~al.}{2007}]{datta07}
{Datta} K.~K.,  {Bharadwaj} S.,   {Choudhury} T.~R.,  2007, \mn@doi [\mnras]
  {10.1111/j.1365-2966.2007.12421.x}, \href
  {http://adsabs.harvard.edu/abs/2007MNRAS.382..809D} {382, 809}

\bibitem[\protect\citeauthoryear{{Datta}, {Majumdar}, {Bharadwaj}  \&
  {Choudhury}}{{Datta} et~al.}{2008}]{datta08}
{Datta} K.~K.,  {Majumdar} S.,  {Bharadwaj} S.,   {Choudhury} T.~R.,  2008,
  \mn@doi [\mnras] {10.1111/j.1365-2966.2008.14008.x}, \href
  {http://adsabs.harvard.edu/abs/2008MNRAS.391.1900D} {391, 1900}

\bibitem[\protect\citeauthoryear{{Datta}, {Friedrich}, {Mellema}, {Iliev}  \&
  {Shapiro}}{{Datta} et~al.}{2012}]{datta12}
{Datta} K.~K.,  {Friedrich} M.~M.,  {Mellema} G.,  {Iliev} I.~T.,   {Shapiro}
  P.~R.,  2012, \mn@doi [\mnras] {10.1111/j.1365-2966.2012.21268.x}, \href
  {http://adsabs.harvard.edu/abs/2012MNRAS.424..762D} {424, 762}

\bibitem[\protect\citeauthoryear{{Datta}, {Ghara}, {Majumdar}, {Choudhury},
  {Bharadwaj}, {Roy}  \& {Datta}}{{Datta} et~al.}{2016}]{Datta16}
{Datta} K.~K.,  {Ghara} R.,  {Majumdar} S.,  {Choudhury} T.~R.,  {Bharadwaj}
  S.,  {Roy} H.,   {Datta} A.,  2016, \mn@doi [Journal of Astrophysics and
  Astronomy] {10.1007/s12036-016-9405-x}, \href
  {http://adsabs.harvard.edu/abs/2016JApA...37...27D} {37, 27}

\bibitem[\protect\citeauthoryear{{Davis}, {Efstathiou}, {Frenk}  \&
  {White}}{{Davis} et~al.}{1985}]{davis85}
{Davis} M.,  {Efstathiou} G.,  {Frenk} C.~S.,   {White} S.~D.~M.,  1985,
  \mn@doi [\apj] {10.1086/163168}, \href
  {http://adsabs.harvard.edu/abs/1985ApJ...292..371D} {292, 371}

\bibitem[\protect\citeauthoryear{{Dayal} \& {Ferrara}}{{Dayal} \&
  {Ferrara}}{2018}]{dayal2018}
{Dayal} P.,  {Ferrara} A.,  2018, \mn@doi [\physrep]
  {10.1016/j.physrep.2018.10.002}, \href
  {http://adsabs.harvard.edu/abs/2018PhR...780....1D} {780, 1}

\bibitem[\protect\citeauthoryear{{Dayal}, {Dunlop}, {Maio}  \&
  {Ciardi}}{{Dayal} et~al.}{2013}]{Dayal13}
{Dayal} P.,  {Dunlop} J.~S.,  {Maio} U.,   {Ciardi} B.,  2013, \mn@doi [\mnras]
  {10.1093/mnras/stt1108}, \href
  {http://adsabs.harvard.edu/abs/2013MNRAS.434.1486D} {434, 1486}

\bibitem[\protect\citeauthoryear{{Evans} et~al.,}{{Evans}
  et~al.}{2015}]{Evans15}
{Evans} C.,  et~al., 2015, preprint, \href
  {http://adsabs.harvard.edu/abs/2015arXiv150104726E} {} (\mn@eprint {arXiv}
  {1501.04726})

\bibitem[\protect\citeauthoryear{{Finlator}, {Oppenheimer}  \&
  {Dav{\'e}}}{{Finlator} et~al.}{2011}]{Finlator11}
{Finlator} K.,  {Oppenheimer} B.~D.,   {Dav{\'e}} R.,  2011, \mn@doi [\mnras]
  {10.1111/j.1365-2966.2010.17554.x}, \href
  {http://adsabs.harvard.edu/abs/2011MNRAS.410.1703F} {410, 1703}

\bibitem[\protect\citeauthoryear{{Furlanetto}, {Zaldarriaga}  \&
  {Hernquist}}{{Furlanetto} et~al.}{2004}]{furlanetto04}
{Furlanetto} S.~R.,  {Zaldarriaga} M.,   {Hernquist} L.,  2004, \mn@doi [\apj]
  {10.1086/423025}, \href {http://adsabs.harvard.edu/abs/2004ApJ...613....1F}
  {613, 1}

\bibitem[\protect\citeauthoryear{{Geil}, {Mutch}, {Poole}, {Duffy}, {Mesinger}
  \& {Wyithe}}{{Geil} et~al.}{2017}]{Geil17}
{Geil} P.~M.,  {Mutch} S.~J.,  {Poole} G.~B.,  {Duffy} A.~R.,  {Mesinger} A.,
  {Wyithe} J.~S.~B.,  2017, preprint, \href
  {http://adsabs.harvard.edu/abs/2017arXiv170405175G} {} (\mn@eprint {arXiv}
  {1704.05175})

\bibitem[\protect\citeauthoryear{{Ghara}, {Choudhury}  \& {Datta}}{{Ghara}
  et~al.}{2016}]{Ghara16}
{Ghara} R.,  {Choudhury} T.~R.,   {Datta} K.~K.,  2016, \mn@doi [\mnras]
  {10.1093/mnras/stw953}, \href
  {http://adsabs.harvard.edu/abs/2016MNRAS.460..827G} {460, 827}

\bibitem[\protect\citeauthoryear{{Giri}, {Mellema}  \& {Ghara}}{{Giri}
  et~al.}{2018}]{Giri18}
{Giri} S.~K.,  {Mellema} G.,   {Ghara} R.,  2018, \mn@doi [\mnras]
  {10.1093/mnras/sty1786}, \href
  {http://adsabs.harvard.edu/abs/2018MNRAS.479.5596G} {479, 5596}

\bibitem[\protect\citeauthoryear{{Greig} \& {Mesinger}}{{Greig} \&
  {Mesinger}}{2017a}]{Greig17a}
{Greig} B.,  {Mesinger} A.,  2017a, \mn@doi [\mnras] {10.1093/mnras/stw3026},
  \href {http://adsabs.harvard.edu/abs/2017MNRAS.465.4838G} {465, 4838}

\bibitem[\protect\citeauthoryear{{Greig} \& {Mesinger}}{{Greig} \&
  {Mesinger}}{2017b}]{Greig17b}
{Greig} B.,  {Mesinger} A.,  2017b, \mn@doi [\mnras] {10.1093/mnras/stx2118},
  \href {https://ui.adsabs.harvard.edu/abs/2017MNRAS.472.2651G} {472, 2651}

\bibitem[\protect\citeauthoryear{{Gr{\"u}tzbauch} et~al.,}{{Gr{\"u}tzbauch}
  et~al.}{2011}]{Grutzbauch11}
{Gr{\"u}tzbauch} R.,  et~al., 2011, \mn@doi [\mnras]
  {10.1111/j.1365-2966.2011.19559.x}, \href
  {http://adsabs.harvard.edu/abs/2011MNRAS.418..938G} {418, 938}

\bibitem[\protect\citeauthoryear{{Harikane} et~al.,}{{Harikane}
  et~al.}{2019}]{Harikane19}
{Harikane} Y.,  et~al., 2019, arXiv e-prints, \href
  {http://adsabs.harvard.edu/abs/2019arXiv190209555H} {}

\bibitem[\protect\citeauthoryear{{Hasegawa} et~al.,}{{Hasegawa}
  et~al.}{2016}]{Hasegawa16}
{Hasegawa} K.,  et~al., 2016, preprint, \href
  {http://adsabs.harvard.edu/abs/2016arXiv160301961H} {} (\mn@eprint {arXiv}
  {1603.01961})

\bibitem[\protect\citeauthoryear{{Hassan}, {Dav{\'e}}, {Finlator}  \&
  {Santos}}{{Hassan} et~al.}{2017}]{Hassan17}
{Hassan} S.,  {Dav{\'e}} R.,  {Finlator} K.,   {Santos} M.~G.,  2017, \mn@doi
  [\mnras] {10.1093/mnras/stx420}, \href
  {http://adsabs.harvard.edu/abs/2017MNRAS.468..122H} {468, 122}

\bibitem[\protect\citeauthoryear{{Hutter}, {Dayal}, {M{\"u}ller}  \&
  {Trott}}{{Hutter} et~al.}{2016}]{Hutter17}
{Hutter} A.,  {Dayal} P.,  {M{\"u}ller} V.,   {Trott} C.,  2016, preprint,
  \href {http://adsabs.harvard.edu/abs/2016arXiv160501734H} {} (\mn@eprint
  {arXiv} {1605.01734})

\bibitem[\protect\citeauthoryear{{Hutter}, {Trott}  \& {Dayal}}{{Hutter}
  et~al.}{2018}]{Hutter18}
{Hutter} A.,  {Trott} C.~M.,   {Dayal} P.,  2018, \mn@doi [\mnras]
  {10.1093/mnrasl/sly115}, \href
  {http://adsabs.harvard.edu/abs/2018MNRAS.479L.129H} {479, L129}

\bibitem[\protect\citeauthoryear{{Inayoshi}, {Li}  \& {Haiman}}{{Inayoshi}
  et~al.}{2018}]{Inayoshi18}
{Inayoshi} K.,  {Li} M.,   {Haiman} Z.,  2018, \mn@doi [\mnras]
  {10.1093/mnras/sty1720}, \href
  {http://adsabs.harvard.edu/abs/2018MNRAS.479.4017I} {479, 4017}

\bibitem[\protect\citeauthoryear{{Inoue}}{{Inoue}}{2010}]{Inoue10}
{Inoue} A.~K.,  2010, \mn@doi [\mnras] {10.1111/j.1365-2966.2009.15730.x},
  \href {http://adsabs.harvard.edu/abs/2010MNRAS.401.1325I} {401, 1325}

\bibitem[\protect\citeauthoryear{{Inoue} et~al.,}{{Inoue}
  et~al.}{2018}]{Inoue18}
{Inoue} A.~K.,  et~al., 2018, \mn@doi [\pasj] {10.1093/pasj/psy048}, \href
  {http://adsabs.harvard.edu/abs/2018PASJ...70...55I} {70, 55}

\bibitem[\protect\citeauthoryear{{Ishigaki}, {Ouchi}  \& {Harikane}}{{Ishigaki}
  et~al.}{2016}]{Ishigaki16}
{Ishigaki} M.,  {Ouchi} M.,   {Harikane} Y.,  2016, \mn@doi [\apj]
  {10.3847/0004-637X/822/1/5}, \href
  {http://esoads.eso.org/abs/2016ApJ...822....5I} {822, 5}

\bibitem[\protect\citeauthoryear{{Jaacks}, {Nagamine}  \& {Choi}}{{Jaacks}
  et~al.}{2012}]{Jaacks12}
{Jaacks} J.,  {Nagamine} K.,   {Choi} J.~H.,  2012, \mn@doi [\mnras]
  {10.1111/j.1365-2966.2012.21989.x}, \href
  {http://adsabs.harvard.edu/abs/2012MNRAS.427..403J} {427, 403}

\bibitem[\protect\citeauthoryear{{Kakiichi}, {Graziani}, {Ciardi}, {Meiksin},
  {Compostella}, {Eide}  \& {Zaroubi}}{{Kakiichi} et~al.}{2017}]{Kakiichi17}
{Kakiichi} K.,  {Graziani} L.,  {Ciardi} B.,  {Meiksin} A.,  {Compostella} M.,
  {Eide} M.~B.,   {Zaroubi} S.,  2017, \mn@doi [\mnras] {10.1093/mnras/stx603},
  \href {http://adsabs.harvard.edu/abs/2017MNRAS.468.3718K} {468, 3718}

\bibitem[\protect\citeauthoryear{{Kaurov} \& {Gnedin}}{{Kaurov} \&
  {Gnedin}}{2015}]{kaurov15}
{Kaurov} A.~A.,  {Gnedin} N.~Y.,  2015, \mn@doi [\apj]
  {10.1088/0004-637X/810/2/154}, \href
  {http://adsabs.harvard.edu/abs/2015ApJ...810..154K} {810, 154}

\bibitem[\protect\citeauthoryear{{Kimm} \& {Cen}}{{Kimm} \&
  {Cen}}{2014}]{Kimm&Cen14}
{Kimm} T.,  {Cen} R.,  2014, \mn@doi [\apj] {10.1088/0004-637X/788/2/121},
  \href {http://adsabs.harvard.edu/abs/2014ApJ...788..121K} {788, 121}

\bibitem[\protect\citeauthoryear{{Koopmans} et~al.,}{{Koopmans}
  et~al.}{2015}]{Koopmans15}
{Koopmans} L.,  et~al., 2015, in Advancing Astrophysics with the Square
  Kilometre Array (AASKA14). p.~1 (\mn@eprint {arXiv} {1505.07568})

\bibitem[\protect\citeauthoryear{{Kroupa}}{{Kroupa}}{2001}]{Kroupa01}
{Kroupa} P.,  2001, \mn@doi [\mnras] {10.1046/j.1365-8711.2001.04022.x}, \href
  {http://adsabs.harvard.edu/abs/2001MNRAS.322..231K} {322, 231}

\bibitem[\protect\citeauthoryear{{Kulkarni}, {Hennawi}, {Rollinde}  \&
  {Vangioni}}{{Kulkarni} et~al.}{2014}]{Kulkarni14}
{Kulkarni} G.,  {Hennawi} J.~F.,  {Rollinde} E.,   {Vangioni} E.,  2014,
  \mn@doi [\apj] {10.1088/0004-637X/787/1/64}, \href
  {http://adsabs.harvard.edu/abs/2014ApJ...787...64K} {787, 64}

\bibitem[\protect\citeauthoryear{{Kulkarni}, {Choudhury}, {Puchwein}  \&
  {Haehnelt}}{{Kulkarni} et~al.}{2016}]{Kulkarni16}
{Kulkarni} G.,  {Choudhury} T.~R.,  {Puchwein} E.,   {Haehnelt} M.~G.,  2016,
  \mn@doi [\mnras] {10.1093/mnras/stw2168}, \href
  {http://adsabs.harvard.edu/abs/2016MNRAS.463.2583K} {463, 2583}

\bibitem[\protect\citeauthoryear{{Laporte} et~al.,}{{Laporte}
  et~al.}{2017}]{Laporte17}
{Laporte} N.,  et~al., 2017, \mn@doi [\apjl] {10.3847/2041-8213/aa62aa}, \href
  {http://esoads.eso.org/abs/2017ApJ...837L..21L} {837, L21}

\bibitem[\protect\citeauthoryear{{Lee}, {Primack}, {Behroozi},
  {Rodr{\'{\i}}guez-Puebla}, {Hellinger}  \& {Dekel}}{{Lee}
  et~al.}{2017}]{Lee17}
{Lee} C.~T.,  {Primack} J.~R.,  {Behroozi} P.,  {Rodr{\'{\i}}guez-Puebla} A.,
  {Hellinger} D.,   {Dekel} A.,  2017, \mn@doi [\mnras]
  {10.1093/mnras/stw3348}, \href
  {http://adsabs.harvard.edu/abs/2017MNRAS.466.3834L} {466, 3834}

\bibitem[\protect\citeauthoryear{{Leitherer} et~al.,}{{Leitherer}
  et~al.}{1999}]{Leitherer99}
{Leitherer} C.,  et~al., 1999, \mn@doi [\apjs] {10.1086/313233}, \href
  {http://adsabs.harvard.edu/abs/1999ApJS..123....3L} {123, 3}

\bibitem[\protect\citeauthoryear{Lewis, Challinor  \& Lasenby}{Lewis
  et~al.}{2000}]{Lewis:1999bs}
Lewis A.,  Challinor A.,   Lasenby A.,  2000, \mn@doi [Astrophys. J.]
  {10.1086/309179}, 538, 473

\bibitem[\protect\citeauthoryear{{Lidz}, {Zahn}, {Furlanetto}, {McQuinn},
  {Hernquist}  \& {Zaldarriaga}}{{Lidz} et~al.}{2009}]{Lidz09}
{Lidz} A.,  {Zahn} O.,  {Furlanetto} S.~R.,  {McQuinn} M.,  {Hernquist} L.,
  {Zaldarriaga} M.,  2009, \mn@doi [\apj] {10.1088/0004-637X/690/1/252}, \href
  {http://adsabs.harvard.edu/abs/2009ApJ...690..252L} {690, 252}

\bibitem[\protect\citeauthoryear{{Livermore}, {Finkelstein}  \&
  {Lotz}}{{Livermore} et~al.}{2017}]{Livermore17}
{Livermore} R.~C.,  {Finkelstein} S.~L.,   {Lotz} J.~M.,  2017, \mn@doi [\apj]
  {10.3847/1538-4357/835/2/113}, \href
  {http://adsabs.harvard.edu/abs/2017ApJ...835..113L} {835, 113}

\bibitem[\protect\citeauthoryear{{Loeb} \& {Furlanetto}}{{Loeb} \&
  {Furlanetto}}{2013}]{loeb13}
{Loeb} A.,  {Furlanetto} S.~R.,  2013, {The First Galaxies in the Universe}

\bibitem[\protect\citeauthoryear{{Ma}, {Kasen}, {Hopkins},
  {Faucher-Gigu{\`e}re}, {Quataert}, {Kere{\v s}}  \& {Murray}}{{Ma}
  et~al.}{2015}]{Ma15}
{Ma} X.,  {Kasen} D.,  {Hopkins} P.~F.,  {Faucher-Gigu{\`e}re} C.-A.,
  {Quataert} E.,  {Kere{\v s}} D.,   {Murray} N.,  2015, \mn@doi [\mnras]
  {10.1093/mnras/stv1679}, \href
  {http://adsabs.harvard.edu/abs/2015MNRAS.453..960M} {453, 960}

\bibitem[\protect\citeauthoryear{{Ma} et~al.,}{{Ma} et~al.}{2018}]{Ma18}
{Ma} X.,  et~al., 2018, \mn@doi [\mnras] {10.1093/mnras/sty1024}, \href
  {http://adsabs.harvard.edu/abs/2018MNRAS.478.1694M} {478, 1694}

\bibitem[\protect\citeauthoryear{{Maio}, {Ciardi}, {Dolag}, {Tornatore}  \&
  {Khochfar}}{{Maio} et~al.}{2010}]{Maio10}
{Maio} U.,  {Ciardi} B.,  {Dolag} K.,  {Tornatore} L.,   {Khochfar} S.,  2010,
  \mn@doi [\mnras] {10.1111/j.1365-2966.2010.17003.x}, \href
  {http://adsabs.harvard.edu/abs/2010MNRAS.407.1003M} {407, 1003}

\bibitem[\protect\citeauthoryear{{Maio}, {Petkova}, {De Lucia}  \&
  {Borgani}}{{Maio} et~al.}{2016}]{Maio16}
{Maio} U.,  {Petkova} M.,  {De Lucia} G.,   {Borgani} S.,  2016, \mn@doi
  [\mnras] {10.1093/mnras/stw1196}, \href
  {http://adsabs.harvard.edu/abs/2016MNRAS.460.3733M} {460, 3733}

\bibitem[\protect\citeauthoryear{{Majumdar}, {Bharadwaj}, {Datta}  \&
  {Choudhury}}{{Majumdar} et~al.}{2011}]{majumdar11}
{Majumdar} S.,  {Bharadwaj} S.,  {Datta} K.~K.,   {Choudhury} T.~R.,  2011,
  \mn@doi [\mnras] {10.1111/j.1365-2966.2011.18223.x}, \href
  {http://adsabs.harvard.edu/abs/2011MNRAS.413.1409M} {413, 1409}

\bibitem[\protect\citeauthoryear{{Majumdar}, {Bharadwaj}  \&
  {Choudhury}}{{Majumdar} et~al.}{2012}]{majumdar12}
{Majumdar} S.,  {Bharadwaj} S.,   {Choudhury} T.~R.,  2012, \mn@doi [\mnras]
  {10.1111/j.1365-2966.2012.21914.x}, \href
  {http://adsabs.harvard.edu/abs/2012MNRAS.426.3178M} {426, 3178}

\bibitem[\protect\citeauthoryear{{Majumdar}, {Mellema}, {Datta}, {Jensen},
  {Choudhury}, {Bharadwaj}  \& {Friedrich}}{{Majumdar}
  et~al.}{2014}]{majumdar14}
{Majumdar} S.,  {Mellema} G.,  {Datta} K.~K.,  {Jensen} H.,  {Choudhury} T.~R.,
   {Bharadwaj} S.,   {Friedrich} M.~M.,  2014, \mn@doi [\mnras]
  {10.1093/mnras/stu1342}, \href
  {http://adsabs.harvard.edu/abs/2014MNRAS.443.2843M} {443, 2843}

\bibitem[\protect\citeauthoryear{{Malloy} \& {Lidz}}{{Malloy} \&
  {Lidz}}{2013}]{malloy13}
{Malloy} M.,  {Lidz} A.,  2013, \mn@doi [\apj] {10.1088/0004-637X/767/1/68},
  \href {http://adsabs.harvard.edu/abs/2013ApJ...767...68M} {767, 68}

\bibitem[\protect\citeauthoryear{{Marchetti}, {Serjeant}  \&
  {Vaccari}}{{Marchetti} et~al.}{2017}]{Marchetti17}
{Marchetti} L.,  {Serjeant} S.,   {Vaccari} M.,  2017, \mn@doi [\mnras]
  {10.1093/mnras/stx1553}, \href
  {http://adsabs.harvard.edu/abs/2017MNRAS.470.5007M} {470, 5007}

\bibitem[\protect\citeauthoryear{{Maselli}, {Gallerani}, {Ferrara}  \&
  {Choudhury}}{{Maselli} et~al.}{2007}]{Maselli07}
{Maselli} A.,  {Gallerani} S.,  {Ferrara} A.,   {Choudhury} T.~R.,  2007,
  \mn@doi [\mnras] {10.1111/j.1745-3933.2007.00283.x}, \href
  {http://adsabs.harvard.edu/abs/2007MNRAS.376L..34M} {376, L34}

\bibitem[\protect\citeauthoryear{{McNaught-Roberts} et~al.,}{{McNaught-Roberts}
  et~al.}{2014}]{McNaught-Roberts14}
{McNaught-Roberts} T.,  et~al., 2014, \mn@doi [\mnras] {10.1093/mnras/stu1886},
  \href {http://adsabs.harvard.edu/abs/2014MNRAS.445.2125M} {445, 2125}

\bibitem[\protect\citeauthoryear{{McQuinn}, {Lidz}, {Zahn}, {Dutta},
  {Hernquist}  \& {Zaldarriaga}}{{McQuinn} et~al.}{2007}]{McQuinn07}
{McQuinn} M.,  {Lidz} A.,  {Zahn} O.,  {Dutta} S.,  {Hernquist} L.,
  {Zaldarriaga} M.,  2007, \mn@doi [\mnras] {10.1111/j.1365-2966.2007.11489.x},
  \href {http://adsabs.harvard.edu/abs/2007MNRAS.377.1043M} {377, 1043}

\bibitem[\protect\citeauthoryear{{Mellema}, {Koopmans}, {Shukla}, {Datta},
  {Mesinger}  \& {Majumdar}}{{Mellema} et~al.}{2015}]{Mellema15}
{Mellema} G.,  {Koopmans} L.,  {Shukla} H.,  {Datta} K.~K.,  {Mesinger} A.,
  {Majumdar} S.,  2015, Advancing Astrophysics with the Square Kilometre Array
  (AASKA14), \href {http://adsabs.harvard.edu/abs/2015aska.confE..10M} {p.~10}

\bibitem[\protect\citeauthoryear{{Mesinger}}{{Mesinger}}{2016}]{Mesinger16}
{Mesinger} A.,  ed. 2016, {Understanding the Epoch of Cosmic Reionization}
  Astrophysics and Space Science Library Vol. 423,
  \mn@doi{10.1007/978-3-319-21957-8.
}

\bibitem[\protect\citeauthoryear{{Mesinger} \& {Dijkstra}}{{Mesinger} \&
  {Dijkstra}}{2008}]{Mesinger08}
{Mesinger} A.,  {Dijkstra} M.,  2008, \mn@doi [\mnras]
  {10.1111/j.1365-2966.2008.13776.x}, \href
  {http://adsabs.harvard.edu/abs/2008MNRAS.390.1071M} {390, 1071}

\bibitem[\protect\citeauthoryear{{Mesinger} \& {Furlanetto}}{{Mesinger} \&
  {Furlanetto}}{2007}]{Mesinger07}
{Mesinger} A.,  {Furlanetto} S.,  2007, \mn@doi [\apj] {10.1086/521806}, \href
  {http://adsabs.harvard.edu/abs/2007ApJ...669..663M} {669, 663}

\bibitem[\protect\citeauthoryear{{Mirocha}, {Furlanetto}  \& {Sun}}{{Mirocha}
  et~al.}{2017}]{Mirocha17}
{Mirocha} J.,  {Furlanetto} S.~R.,   {Sun} G.,  2017, \mn@doi [\mnras]
  {10.1093/mnras/stw2412}, \href
  {http://adsabs.harvard.edu/abs/2017MNRAS.464.1365M} {464, 1365}

\bibitem[\protect\citeauthoryear{{Mondal}, {Bharadwaj}, {Majumdar}, {Bera}  \&
  {Acharyya}}{{Mondal} et~al.}{2015}]{Mondal15}
{Mondal} R.,  {Bharadwaj} S.,  {Majumdar} S.,  {Bera} A.,   {Acharyya} A.,
  2015, \mn@doi [\mnras] {10.1093/mnrasl/slv015}, \href
  {http://adsabs.harvard.edu/abs/2015MNRAS.449L..41M} {449, L41}

\bibitem[\protect\citeauthoryear{{Mondal}, {Bharadwaj}  \& {Majumdar}}{{Mondal}
  et~al.}{2016}]{mondal16}
{Mondal} R.,  {Bharadwaj} S.,   {Majumdar} S.,  2016, \mn@doi [\mnras]
  {10.1093/mnras/stv2772}, \href
  {http://adsabs.harvard.edu/abs/2016MNRAS.456.1936M} {456, 1936}

\bibitem[\protect\citeauthoryear{{Mondal}, {Bharadwaj}  \& {Majumdar}}{{Mondal}
  et~al.}{2017}]{Mondal17}
{Mondal} R.,  {Bharadwaj} S.,   {Majumdar} S.,  2017, \mn@doi [\mnras]
  {10.1093/mnras/stw2599}, \href
  {http://adsabs.harvard.edu/abs/2017MNRAS.464.2992M} {464, 2992}

\bibitem[\protect\citeauthoryear{{Mondal}, {Bharadwaj}  \& {Datta}}{{Mondal}
  et~al.}{2018}]{Mondal18}
{Mondal} R.,  {Bharadwaj} S.,   {Datta} K.~K.,  2018, \mn@doi [\mnras]
  {10.1093/mnras/stx2888}, \href
  {http://adsabs.harvard.edu/abs/2018MNRAS.474.1390M} {474, 1390}

\bibitem[\protect\citeauthoryear{{Mondal}, {Bharadwaj}, {Iliev}, {Datta},
  {Majumdar}, {Shaw}  \& {Sarkar}}{{Mondal} et~al.}{2019}]{Mondal18a}
{Mondal} R.,  {Bharadwaj} S.,  {Iliev} I.~T.,  {Datta} K.~K.,  {Majumdar} S.,
  {Shaw} A.~K.,   {Sarkar} A.~K.,  2019, \mn@doi [\mnras]
  {10.1093/mnrasl/sly226}, \href
  {http://adsabs.harvard.edu/abs/2019MNRAS.483L.109M} {483, L109}

\bibitem[\protect\citeauthoryear{{Monsalve}, {Greig}, {Bowman}, {Mesinger},
  {Rogers}, {Mozdzen}, {Kern}  \& {Mahesh}}{{Monsalve}
  et~al.}{2018}]{Monsalve18}
{Monsalve} R.~A.,  {Greig} B.,  {Bowman} J.~D.,  {Mesinger} A.,  {Rogers} A.
  E.~E.,  {Mozdzen} T.~J.,  {Kern} N.~S.,   {Mahesh} N.,  2018, \mn@doi [\apj]
  {10.3847/1538-4357/aace54}, \href
  {https://ui.adsabs.harvard.edu/abs/2018ApJ...863...11M} {863, 11}

\bibitem[\protect\citeauthoryear{{Mutch}, {Geil}, {Poole}, {Angel}, {Duffy},
  {Mesinger}  \& {Wyithe}}{{Mutch} et~al.}{2016}]{Mutch16}
{Mutch} S.~J.,  {Geil} P.~M.,  {Poole} G.~B.,  {Angel} P.~W.,  {Duffy} A.~R.,
  {Mesinger} A.,   {Wyithe} J.~S.~B.,  2016, \mn@doi [\mnras]
  {10.1093/mnras/stw1506}, \href
  {http://adsabs.harvard.edu/abs/2016MNRAS.462..250M} {462, 250}

\bibitem[\protect\citeauthoryear{{Overzier}}{{Overzier}}{2016}]{Overzier16}
{Overzier} R.~A.,  2016, \mn@doi [\aapr] {10.1007/s00159-016-0100-3}, \href
  {http://esoads.eso.org/abs/2016A%26ARv..24...14O} {24, 14}

\bibitem[\protect\citeauthoryear{{Paardekooper}, {Khochfar}  \& {Dalla
  Vecchia}}{{Paardekooper} et~al.}{2015}]{Paardekooper15}
{Paardekooper} J.-P.,  {Khochfar} S.,   {Dalla Vecchia} C.,  2015, \mn@doi
  [\mnras] {10.1093/mnras/stv1114}, \href
  {http://adsabs.harvard.edu/abs/2015MNRAS.451.2544P} {451, 2544}

\bibitem[\protect\citeauthoryear{{Pacucci}, {Mesinger}, {Mineo}  \&
  {Ferrara}}{{Pacucci} et~al.}{2014}]{Pacucci14}
{Pacucci} F.,  {Mesinger} A.,  {Mineo} S.,   {Ferrara} A.,  2014, \mn@doi
  [\mnras] {10.1093/mnras/stu1240}, \href
  {http://adsabs.harvard.edu/abs/2014MNRAS.443..678P} {443, 678}

\bibitem[\protect\citeauthoryear{{Park}, {Kim}, {Wyithe}  \& {Lacey}}{{Park}
  et~al.}{2014}]{Park14}
{Park} J.,  {Kim} H.-S.,  {Wyithe} J.~S.~B.,   {Lacey} C.~G.,  2014, \mn@doi
  [\mnras] {10.1093/mnras/stt2366}, \href
  {http://adsabs.harvard.edu/abs/2014MNRAS.438.2474P} {438, 2474}

\bibitem[\protect\citeauthoryear{{Park}, {Shapiro}, {Choi}, {Yoshida}, {Hirano}
   \& {Ahn}}{{Park} et~al.}{2016}]{park16}
{Park} H.,  {Shapiro} P.~R.,  {Choi} J.-h.,  {Yoshida} N.,  {Hirano} S.,
  {Ahn} K.,  2016, \mn@doi [\apj] {10.3847/0004-637X/831/1/86}, \href
  {http://adsabs.harvard.edu/abs/2016ApJ...831...86P} {831, 86}

\bibitem[\protect\citeauthoryear{{Planck Collaboration} et~al.,}{{Planck
  Collaboration} et~al.}{2014}]{Planck14}
{Planck Collaboration} et~al., 2014, \mn@doi [\aap]
  {10.1051/0004-6361/201321591}, \href
  {http://adsabs.harvard.edu/abs/2014A%26A...571A..16P} {571, A16}

\bibitem[\protect\citeauthoryear{{Planck Collaboration} et~al.,}{{Planck
  Collaboration} et~al.}{2016}]{planck16a}
{Planck Collaboration} et~al., 2016, \mn@doi [\aap]
  {10.1051/0004-6361/201628897}, \href
  {http://adsabs.harvard.edu/abs/2016A%26A...596A.108P} {596, A108}

\bibitem[\protect\citeauthoryear{{Rodr{\'{\i}}guez-Puebla}, {Behroozi},
  {Primack}, {Klypin}, {Lee}  \& {Hellinger}}{{Rodr{\'{\i}}guez-Puebla}
  et~al.}{2016}]{2016MNRAS.462..893R}
{Rodr{\'{\i}}guez-Puebla} A.,  {Behroozi} P.,  {Primack} J.,  {Klypin} A.,
  {Lee} C.,   {Hellinger} D.,  2016, \mn@doi [\mnras] {10.1093/mnras/stw1705},
  \href {http://adsabs.harvard.edu/abs/2016MNRAS.462..893R} {462, 893}

\bibitem[\protect\citeauthoryear{{Sadoun}, {Shlosman}, {Choi}  \&
  {Romano-D{\'{\i}}az}}{{Sadoun} et~al.}{2016}]{Sadoun16}
{Sadoun} R.,  {Shlosman} I.,  {Choi} J.-H.,   {Romano-D{\'{\i}}az} E.,  2016,
  \mn@doi [\apj] {10.3847/0004-637X/829/2/71}, \href
  {http://adsabs.harvard.edu/abs/2016ApJ...829...71S} {829, 71}

\bibitem[\protect\citeauthoryear{{Sahl{\'e}n} et~al.,}{{Sahl{\'e}n}
  et~al.}{2009}]{Sahlen09}
{Sahl{\'e}n} M.,  et~al., 2009, \mn@doi [\mnras]
  {10.1111/j.1365-2966.2009.14923.x}, \href
  {http://adsabs.harvard.edu/abs/2009MNRAS.397..577S} {397, 577}

\bibitem[\protect\citeauthoryear{{Schaerer}}{{Schaerer}}{2002}]{Schaerer02}
{Schaerer} D.,  2002, \mn@doi [\aap] {10.1051/0004-6361:20011619}, \href
  {http://adsabs.harvard.edu/abs/2002A%26A...382...28S} {382, 28}

\bibitem[\protect\citeauthoryear{{Schaye}}{{Schaye}}{2001}]{schaye01}
{Schaye} J.,  2001, \mn@doi [\apj] {10.1086/322421}, \href
  {http://adsabs.harvard.edu/abs/2001ApJ...559..507S} {559, 507}

\bibitem[\protect\citeauthoryear{{Sharma}, {Theuns}, {Frenk}, {Bower}, {Crain},
  {Schaller}  \& {Schaye}}{{Sharma} et~al.}{2017}]{Sharma17}
{Sharma} M.,  {Theuns} T.,  {Frenk} C.,  {Bower} R.~G.,  {Crain} R.~A.,
  {Schaller} M.,   {Schaye} J.,  2017, \mn@doi [\mnras] {10.1093/mnras/stx578},
  \href {http://adsabs.harvard.edu/abs/2017MNRAS.468.2176S} {468, 2176}

\bibitem[\protect\citeauthoryear{{Shimizu}, {Inoue}, {Okamoto}  \&
  {Yoshida}}{{Shimizu} et~al.}{2014}]{Shimizu14}
{Shimizu} I.,  {Inoue} A.~K.,  {Okamoto} T.,   {Yoshida} N.,  2014, \mn@doi
  [\mnras] {10.1093/mnras/stu265}, \href
  {http://adsabs.harvard.edu/abs/2014MNRAS.440..731S} {440, 731}

\bibitem[\protect\citeauthoryear{{Shimizu}, {Inoue}, {Okamoto}  \&
  {Yoshida}}{{Shimizu} et~al.}{2016}]{Shimizu16}
{Shimizu} I.,  {Inoue} A.~K.,  {Okamoto} T.,   {Yoshida} N.,  2016, \mn@doi
  [\mnras] {10.1093/mnras/stw1423}, \href
  {http://adsabs.harvard.edu/abs/2016MNRAS.461.3563S} {461, 3563}

\bibitem[\protect\citeauthoryear{{Sobacchi} \& {Mesinger}}{{Sobacchi} \&
  {Mesinger}}{2013}]{Sobacchi13}
{Sobacchi} E.,  {Mesinger} A.,  2013, \mn@doi [\mnras] {10.1093/mnras/stt693},
  \href {http://adsabs.harvard.edu/abs/2013MNRAS.432.3340S} {432, 3340}

\bibitem[\protect\citeauthoryear{{Sobacchi} \& {Mesinger}}{{Sobacchi} \&
  {Mesinger}}{2014}]{Sobacchi14}
{Sobacchi} E.,  {Mesinger} A.,  2014, \mn@doi [\mnras] {10.1093/mnras/stu377},
  \href {http://adsabs.harvard.edu/abs/2014MNRAS.440.1662S} {440, 1662}

\bibitem[\protect\citeauthoryear{{Sobacchi}, {Mesinger}  \& {Greig}}{{Sobacchi}
  et~al.}{2016}]{Sobacchi16}
{Sobacchi} E.,  {Mesinger} A.,   {Greig} B.,  2016, \mn@doi [\mnras]
  {10.1093/mnras/stw811}, \href
  {http://adsabs.harvard.edu/abs/2016MNRAS.459.2741S} {459, 2741}

\bibitem[\protect\citeauthoryear{{Songaila} \& {Cowie}}{{Songaila} \&
  {Cowie}}{2010}]{songaila10}
{Songaila} A.,  {Cowie} L.~L.,  2010, \mn@doi [\apj]
  {10.1088/0004-637X/721/2/1448}, \href
  {http://adsabs.harvard.edu/abs/2010ApJ...721.1448S} {721, 1448}

\bibitem[\protect\citeauthoryear{{Stefanon} et~al.,}{{Stefanon}
  et~al.}{2019}]{Stefanon19}
{Stefanon} M.,  et~al., 2019, arXiv e-prints, \href
  {http://adsabs.harvard.edu/abs/2019arXiv190210713S} {}

\bibitem[\protect\citeauthoryear{{Stiavelli} \& {Trenti}}{{Stiavelli} \&
  {Trenti}}{2010}]{Stiavelli10}
{Stiavelli} M.,  {Trenti} M.,  2010, \mn@doi [\apjl]
  {10.1088/2041-8205/716/2/L190}, \href
  {http://adsabs.harvard.edu/abs/2010ApJ...716L.190S} {716, L190}

\bibitem[\protect\citeauthoryear{{Susa}}{{Susa}}{2008}]{Susa08}
{Susa} H.,  2008, \mn@doi [\apj] {10.1086/589964}, \href
  {http://adsabs.harvard.edu/abs/2008ApJ...684..226S} {684, 226}

\bibitem[\protect\citeauthoryear{{Targett}, {Dunlop}  \& {McLure}}{{Targett}
  et~al.}{2012}]{Targett12}
{Targett} T.~A.,  {Dunlop} J.~S.,   {McLure} R.~J.,  2012, \mn@doi [\mnras]
  {10.1111/j.1365-2966.2011.20286.x}, \href
  {http://adsabs.harvard.edu/abs/2012MNRAS.420.3621T} {420, 3621}

\bibitem[\protect\citeauthoryear{{Tozzi}, {Madau}, {Meiksin}  \&
  {Rees}}{{Tozzi} et~al.}{2000}]{Tozzi00}
{Tozzi} P.,  {Madau} P.,  {Meiksin} A.,   {Rees} M.~J.,  2000, \mn@doi [\apj]
  {10.1086/308196}, \href {http://adsabs.harvard.edu/abs/2000ApJ...528..597T}
  {528, 597}

\bibitem[\protect\citeauthoryear{{Vrbanec} et~al.,}{{Vrbanec}
  et~al.}{2016}]{Vrbanec16}
{Vrbanec} D.,  et~al., 2016, \mn@doi [\mnras] {10.1093/mnras/stv2993}, \href
  {http://adsabs.harvard.edu/abs/2016MNRAS.457..666V} {457, 666}

\bibitem[\protect\citeauthoryear{{Wiersma} et~al.,}{{Wiersma}
  et~al.}{2013}]{Wiersma13}
{Wiersma} R.~P.~C.,  et~al., 2013, \mn@doi [\mnras] {10.1093/mnras/stt624},
  \href {http://adsabs.harvard.edu/abs/2013MNRAS.432.2615W} {432, 2615}

\bibitem[\protect\citeauthoryear{{Wood} \& {Loeb}}{{Wood} \&
  {Loeb}}{2000}]{Wood&Loeb}
{Wood} K.,  {Loeb} A.,  2000, \mn@doi [\apj] {10.1086/317775}, \href
  {http://adsabs.harvard.edu/abs/2000ApJ...545...86W} {545, 86}

\bibitem[\protect\citeauthoryear{{Wyithe} \& {Loeb}}{{Wyithe} \&
  {Loeb}}{2007a}]{Wyithe07}
{Wyithe} J.~S.~B.,  {Loeb} A.,  2007a, \mn@doi [\mnras]
  {10.1111/j.1365-2966.2006.11201.x}, \href
  {http://adsabs.harvard.edu/abs/2007MNRAS.374..960W} {374, 960}

\bibitem[\protect\citeauthoryear{{Wyithe} \& {Loeb}}{{Wyithe} \&
  {Loeb}}{2007b}]{WyitheLoeb07}
{Wyithe} J.~S.~B.,  {Loeb} A.,  2007b, \mn@doi [\mnras]
  {10.1111/j.1365-2966.2007.11366.x}, \href
  {http://adsabs.harvard.edu/abs/2007MNRAS.375.1034W} {375, 1034}

\bibitem[\protect\citeauthoryear{{Wyithe}, {Loeb}  \& {Barnes}}{{Wyithe}
  et~al.}{2005}]{wyithe05}
{Wyithe} J. S.~B.,  {Loeb} A.,   {Barnes} D.~G.,  2005, \mn@doi [\apj]
  {10.1086/497160}, \href
  {https://ui.adsabs.harvard.edu/\#abs/2005ApJ...634..715W} {634, 715}

\bibitem[\protect\citeauthoryear{{Wyithe}, {Geil}  \& {Kim}}{{Wyithe}
  et~al.}{2015}]{Wyithe15}
{Wyithe} S.,  {Geil} P.,   {Kim} H.,  2015, Advancing Astrophysics with the
  Square Kilometre Array (AASKA14), \href
  {http://adsabs.harvard.edu/abs/2015aska.confE..15W} {p.~15}

\bibitem[\protect\citeauthoryear{{Xu}, {Wise}, {Norman}, {Ahn}  \&
  {O'Shea}}{{Xu} et~al.}{2016}]{Xu16}
{Xu} H.,  {Wise} J.~H.,  {Norman} M.~L.,  {Ahn} K.,   {O'Shea} B.~W.,  2016,
  \mn@doi [\apj] {10.3847/1538-4357/833/1/84}, \href
  {http://adsabs.harvard.edu/abs/2016ApJ...833...84X} {833, 84}

\bibitem[\protect\citeauthoryear{{Yajima} \& {Khochfar}}{{Yajima} \&
  {Khochfar}}{2017}]{Yajima17b}
{Yajima} H.,  {Khochfar} S.,  2017, \mn@doi [\mnras] {10.1093/mnrasl/slw249},
  \href {http://adsabs.harvard.edu/abs/2017MNRAS.467L..51Y} {467, L51}

\bibitem[\protect\citeauthoryear{{Yajima}, {Choi}  \& {Nagamine}}{{Yajima}
  et~al.}{2011}]{Yajima11}
{Yajima} H.,  {Choi} J.-H.,   {Nagamine} K.,  2011, \mn@doi [\mnras]
  {10.1111/j.1365-2966.2010.17920.x}, \href
  {http://adsabs.harvard.edu/abs/2011MNRAS.412..411Y} {412, 411}

\bibitem[\protect\citeauthoryear{{Yajima}, {Shlosman}, {Romano-D{\'{\i}}az}  \&
  {Nagamine}}{{Yajima} et~al.}{2015}]{Yajima15}
{Yajima} H.,  {Shlosman} I.,  {Romano-D{\'{\i}}az} E.,   {Nagamine} K.,  2015,
  \mn@doi [\mnras] {10.1093/mnras/stv974}, \href
  {http://adsabs.harvard.edu/abs/2015MNRAS.451..418Y} {451, 418}

\bibitem[\protect\citeauthoryear{{Yajima}, {Sugimura}  \& {Hasegawa}}{{Yajima}
  et~al.}{2017}]{Yajima17}
{Yajima} H.,  {Sugimura} K.,   {Hasegawa} K.,  2017, preprint, \href
  {http://adsabs.harvard.edu/abs/2017arXiv170105571Y} {} (\mn@eprint {arXiv}
  {1701.05571})

\bibitem[\protect\citeauthoryear{{Zackrisson}, {Rydberg}, {Schaerer},
  {{\"O}stlin}  \& {Tuli}}{{Zackrisson} et~al.}{2011}]{Zackrisson11}
{Zackrisson} E.,  {Rydberg} C.-E.,  {Schaerer} D.,  {{\"O}stlin} G.,   {Tuli}
  M.,  2011, \mn@doi [\apj] {10.1088/0004-637X/740/1/13}, \href
  {http://adsabs.harvard.edu/abs/2011ApJ...740...13Z} {740, 13}

\bibitem[\protect\citeauthoryear{{Zackrisson} et~al.,}{{Zackrisson}
  et~al.}{2017}]{Zackrisson17}
{Zackrisson} E.,  et~al., 2017, \mn@doi [\apj] {10.3847/1538-4357/836/1/78},
  \href {http://adsabs.harvard.edu/abs/2017ApJ...836...78Z} {836, 78}

\bibitem[\protect\citeauthoryear{{Zahn}, {Lidz}, {McQuinn}, {Dutta},
  {Hernquist}, {Zaldarriaga}  \& {Furlanetto}}{{Zahn} et~al.}{2007}]{Zahn07}
{Zahn} O.,  {Lidz} A.,  {McQuinn} M.,  {Dutta} S.,  {Hernquist} L.,
  {Zaldarriaga} M.,   {Furlanetto} S.~R.,  2007, \mn@doi [\apj]
  {10.1086/509597}, \href {http://adsabs.harvard.edu/abs/2007ApJ...654...12Z}
  {654, 12}

\makeatother
\end{thebibliography}



\appendix

\section{The $N_\mathrm{ion}/L_\mathrm{UV}$ parameter and its relation to the prior star formation history}
\label{Appendix_A}
\begin{figure*}
\includegraphics[width=\columnwidth]{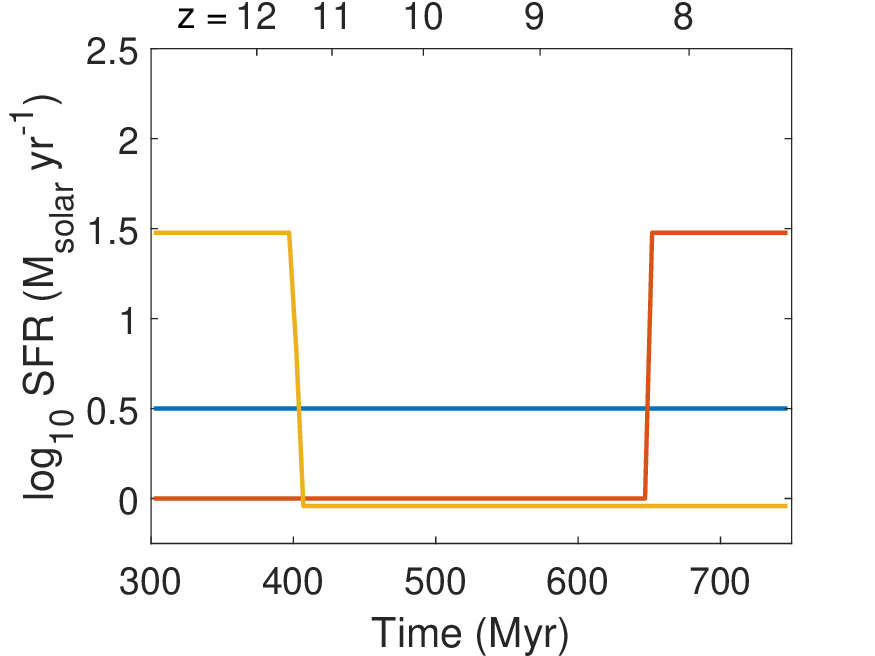}
\includegraphics[width=\columnwidth]{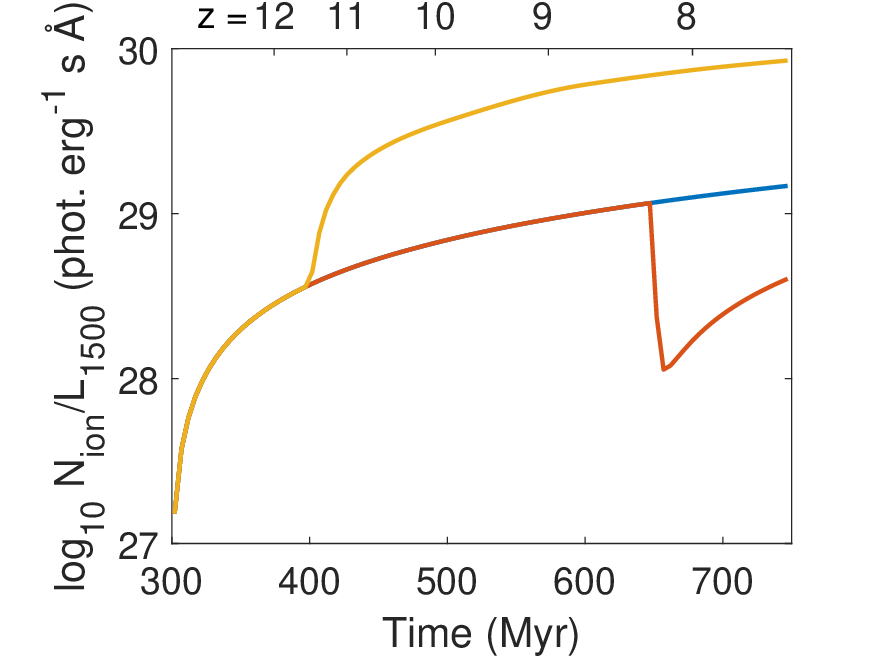}
\caption{Demonstration of how three parametric star formation scenarios for $z\geq 7$ galaxies translate into different predictions on the evolution of the $N_\mathrm{ion}/L_\mathrm{UV}$ ratio. The time axis indicates the age of the Universe from 300 Myr ($z\approx 14$, when star formation is assumed to start) to 750 Myr ($z\approx 7$). {\bf Left:} Star formation histories of the three models. The yellow represents a galaxy where the SFR drops by a factor of $\approx 30$ after 100 Myr, the orange line a galaxy where the SFR remains constant for 350 Myr and then rises by a factor of $\approx 30$ and the blue line a galaxy where the SFR remains constant throughout 450 Myr. {\bf Right:} Prediction of how the $N_\mathrm{ion}/L_\mathrm{1500}$ ratio (cumulative number of ionizing photons produce divided by momentary UV 1500 \AA{} luminosity) evolves for the three star formation scenarios of the left panel. In the constant SFR scenario, the $N_\mathrm{ion,i}/L_\mathrm{1500}$ gradually increases, but drops in SFR (yellow line) leads to higher ratios whereas boosts in SFR (orange line) leads to lower ratios. In these models, a constant metallicity of $Z=0.004$ is assumed and dust attenuation of $L_\mathrm{1500}$ is neglected.}
\label{SFR_Nion_schematic}
\end{figure*}

\begin{figure*}
\includegraphics[width=\columnwidth]{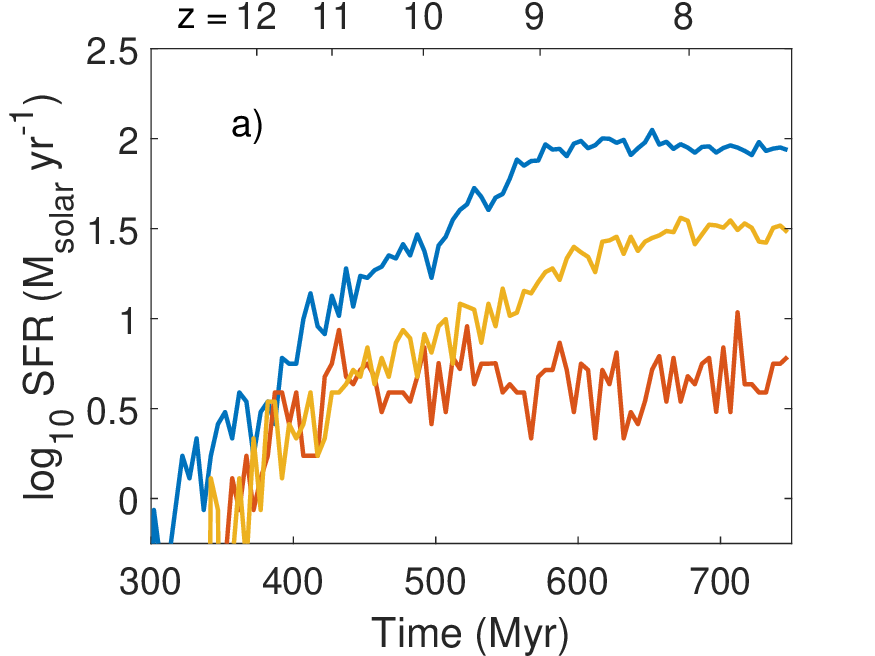}
\includegraphics[width=\columnwidth]{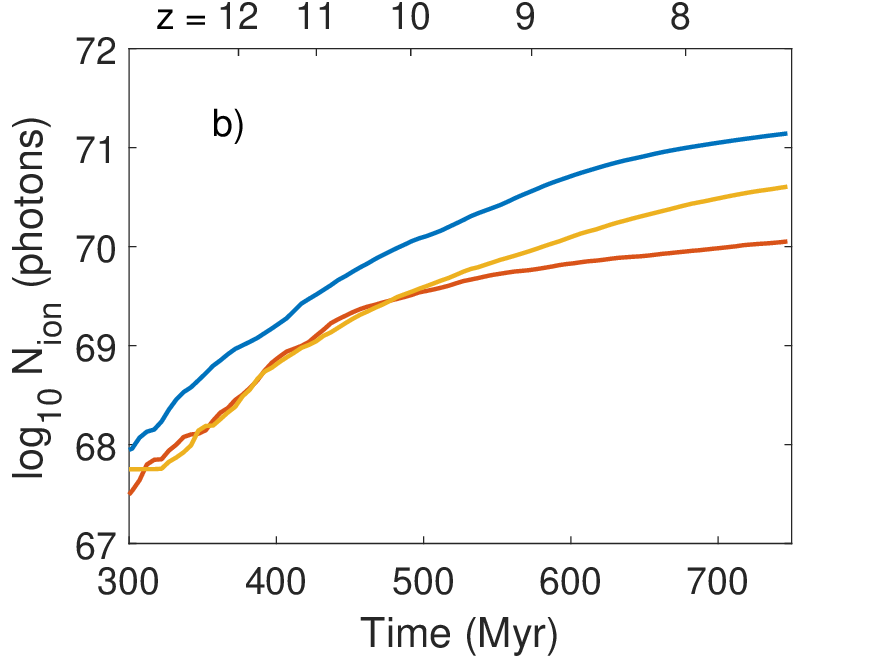}\\
\includegraphics[width=\columnwidth]{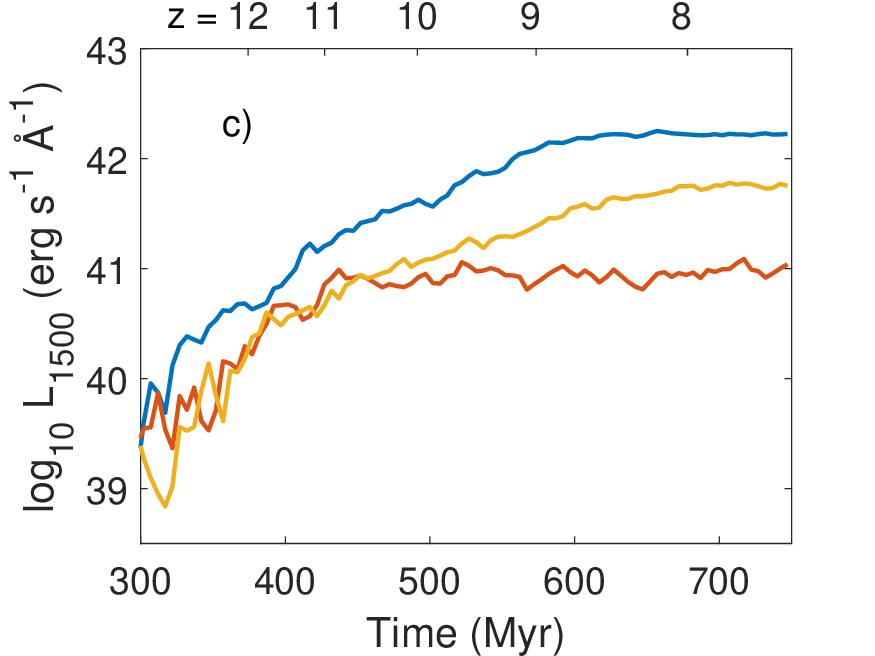}
\includegraphics[width=\columnwidth]{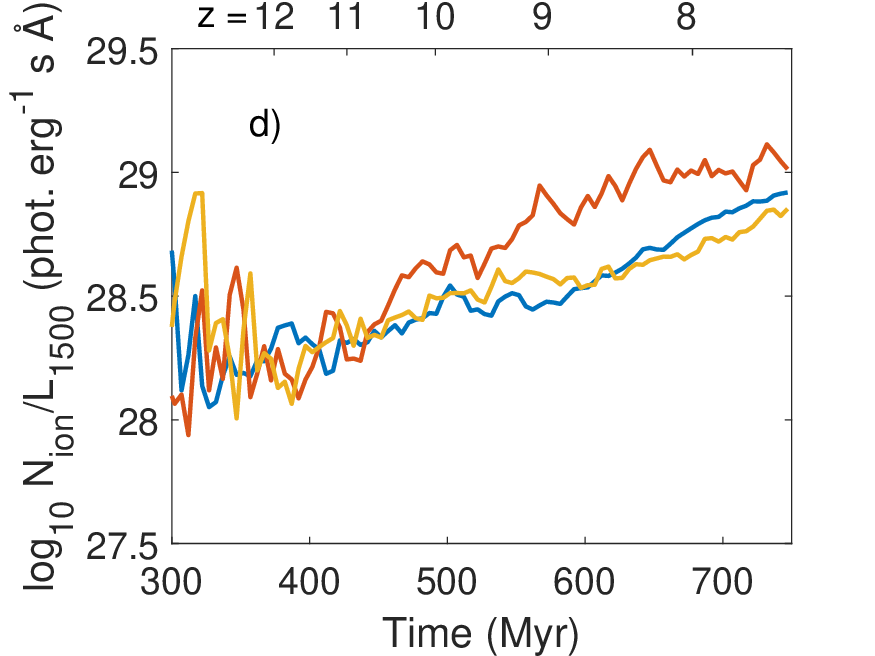}
\caption{Connection between the prior production of ionizing photons and the momentary rest-frame UV luminosity for three simulated high-mass (total stellar mass $\gtrsim 10^9 \ M_\odot$) galaxies at $z \approx 7$. {\bf a)} Star formation activity until 750 Myr after the Big Bang, showing rapid growth followed with fluctuations on short time scales followed by stabilization. {\bf b)} The corresponding cumulative production of ionizing photons {\bf c)} The momentary rest-frame 1500 \AA{} luminosity in the absence of dust effects  {\bf d)} The ratio between the cumulative number of ionizing photons produced and the momentary 1500 \AA{} luminosity, showing fluctuations of factors of only factors a few at late times.}
\label{SFR_Nion_example}
\end{figure*}

In Figure~\ref{SFR_Nion_schematic}, we illustrate the evolution of the $N_\mathrm{ion}/L_\mathrm{UV}$ ratio for three different parametric star formation histories over a few hundred Myr (the time scale relevant for $z>7$ galaxies. A constant SFR leads to a rapid rise in $N_\mathrm{ion}/L_\mathrm{UV}$ during the first 100 Myr after the onset of star formation, followed by a slower increase in this ratio thereafter. Temporal variations in the SFR throughout the star formation history makes $N_\mathrm{ion}/L_\mathrm{UV}$ ratio either go up (following a drop in SFR) or down (for a boost in SFR). 
Galaxies for which the SFR history shows similar trends (e.g. increasing or semi-constant SFR) can be expected to exhibit similar $N_\mathrm{ion}/L_\mathrm{UV}$ ratios. This is demonstrated in Figure~\ref{SFR_Nion_example} using the non-parametric star formation histories of three high-mass galaxies (reaching total stellar mass $\sim 10^9$--$10^{10}\ M_\odot$ by $z\approx 7$) from the \citet{Shimizu16} simulations. These three galaxies all exhibit star formation rates that initially increase over time (albeit with stochastic fluctuations along the way) and then stabilize at different levels of near-constant SFR (Figure~\ref{SFR_Nion_example}a). In the absence of temporal variations in the stellar initial mass function, the number of ionizing photons (or alternatively the total stellar mass) produced becomes a smoothly increasing function of time (Figure~\ref{SFR_Nion_example}b). The rest-frame 1500 \AA{} luminosity (Figure~\ref{SFR_Nion_example}c) traces the SFR fairly faithfully, and the result is a $N_\mathrm{ion}/L_\mathrm{UV}$ ratio (Figure~\ref{SFR_Nion_example}d) that varies no more than a factor of a few after the large initial star formation rate fluctuations that occur at early times.
Please note that the star formation rate histories in these cases includes the star formation that has taken place in-situ within these galaxies, but also within the progenitors that merged to form them. 

To generate Figure~\ref{SFR_Nion_schematic} and Figure~\ref{SFR_Nion_example}, we have for simplicity assumed a dust-free stellar population with constant metallicity $Z=0.004$, and stellar populations SEDs from the Starburst99 stellar population models \citep{Leitherer99}. These results admittedly neglect the effects of dust attenuation on the UV flux and the metallicity distribution within galaxies. Moreover, the star formation history is expected to become increasingly stochastic at lower galaxy masses due to feedback effects \citep[e.g.][]{Shimizu14,Ma18}, which will result in longer spells of low star formation activity and larger galaxy-to-galaxy variations in the $N_\mathrm{ion}/L_\mathrm{UV}$ ratio as a result. All of this is, however, taken into account in the LYCAN machinery for predicting high-redshift galaxy spectra \citep{Zackrisson17} that is used in Sect~\ref{galaxies} of this paper.

\section{Bubble galaxies versus line-of-sight interlopers in photometric surveys}
\label{Appendix_B}
In Sect.~\ref{spec_vs_phot}, we use an argument based on luminosity functions to argue that photometric surveys in the direction of ionized IGM bubbles are less likely to pick up line-of-sight interlopers for bubbles at $z\approx 10$ compared to $z\approx 7$. However, one may also reach the same conclusion by directly comparing the number of high-mass halos present within simulated ionized bubbles, like the one presented in Section~\ref{sizes}, to those in the ambient field along a column spanning $\Delta(z)\approx 1$. In this case, the outcome will depend on the detailed recipe for relating ionzing photon fluxes to halos of a given mass. In the simulations discussed in Section~\ref{sizes}, bubbles with volume $\approx 1000$ cMpc$^3$ contain $\approx 20$--50 halos with virial mass $\geq 10^{10}\ M_\odot/h$ at $z\approx 7$--10. The ambient field will on the other hand contain $\approx 600$ such halos in the $\Delta(z)\approx 1$ volume along the line of sight at $z=7$, yet only $\approx 10$ at $z=10$. This lends further support to the notion that the fraction of line-of-sight interlopers may be very high in photometric bubble surveys at $z\approx 7$, but much smaller at $z\approx 10$. This calculation of the ambient halo density is based on the halo mass function derived from the Bolshoi-Planck and MultiDark-Planck $N$-body simulation \citep{2016MNRAS.462..893R}, and assuming a fiducial flat $\Lambda$CDM cosmology with parameter values $\Omega_{\rm m} = 0.3, \Omega_{\rm b} = 0.044, h = 0.7, \sigma_8=0.8, n_{\rm s}=0.96$. The cosmological computations are performed using a modified version of CAMB \citep{Lewis:1999bs, Sahlen09}.


\bsp	
\label{lastpage}
\end{document}